\newif\ifggtiop
\IfFileExists{iopart.cls}{\ggtioptrue}{\ggtiopfalse}
\ifggtiop
  \documentclass[12pt,final,a4paper]{iopart}
  \usepackage{iopams}
\else
  \documentclass[12pt,a4paper,fleqn]{article}
  \usepackage[margin=24mm]{geometry}
  \usepackage{amsmath,amssymb}
  \newcommand{\fl}{}
  \newcommand{\sref}[1]{section~\ref{#1}}
  
  \newcommand{\address}[1]{\par\begingroup\small\noindent#1\par\endgroup}
  \newcommand{\ead}[1]{\par\begingroup\small\noindent\texttt{#1}\par\endgroup\medskip}
  \newcommand{\pacs}[1]{\par\noindent PACS numbers #1\par\medskip}
  \makeatletter
  \renewcommand{\title}[2][]{\par\noindent{\Large\bfseries#2\par}\medskip}
  \renewcommand{\author}[1]{\par\noindent{\large#1\par}\medskip}
  \makeatother
\fi
\usepackage{amsopn}
\usepackage{mathbbol}
\usepackage{tensor}
\usepackage[makeroom]{cancel}
\usepackage{times}
\usepackage{graphicx}
\usepackage{microtype}
\providecommand{\rmd}{\mathrm{d}}
\providecommand{\rmi}{\mathrm{i}}
\providecommand{\bpi}{\pmb{\pi}}
\providecommand{\bPi}{\pmb{\Pi}}
\providecommand{\bphi}{\pmb{\phi}}
\providecommand{\bPhi}{\pmb{\Phi}}
\providecommand{\Eins}{\mathbb{1}}
\providecommand{\thickspace}{\mskip\thickmuskip}
\newcommand{\onehalf}{{\textstyle\frac{1}{2}}}

\newcommand{\HC}{\mathcal{H}}
\newcommand{\FC}{\tilde{\mathcal{F}}}
\newcommand{\LC}{\mathcal{L}}
\newcommand{\LT}{\tilde{\mathcal{T}}}
\newcommand{\RB}{\mathbb{R}}

\newcommand{\bx}{x}
\newcommand{\by}{y}
\newcommand{\bp}{\pmb{p}}
\newcommand{\bt}{\pmb{t}}
\newcommand{\ba}{\pmb{a}}

\newcommand{\bP}{\pmb{P}}
\newcommand{\bX}{X}
\newcommand{\bFC}{\pmb{\FC}}
\newcommand{\pfrac}[2]{\frac{\partial #1}{\partial #2}}
\newcommand{\ppfrac}[3]{\frac{\partial^2 #1}{\partial #2\partial #3}}
\newcommand{\pppfrac}[4]{\frac{\partial^3 #1}{\partial #2\partial #3\partial #4}}
\newcommand{\fdet}[2]{\pfrac{\big(#1^0,\ldots,#1^3\big)}{\big(#2^0,\ldots,#2^3\big)}}
\newcommand{\detfrac}[2]{\left|\pfrac{#1}{#2}\right|}
\newcommand{\Hexpl}[2]{\left.\pfrac{#1}{#2}\right|_{\mathrm{expl}}}
\begin{document}
\title[Generalized U$(N)$ gauge theory]
{Generalized U$(N)$ gauge transformations in the realm of the extended covariant Hamilton formalism of field theory}
\author{J Struckmeier}
\address{Frankfurt Institute for Advanced Studies (FIAS),\newline
Ruth-Moufang-Str.~1, 60438~Frankfurt am Main, Germany}
\address{Goethe University,
Max-von-Laue-Str.~1, D-60438~Frankfurt am Main, Germany}
\ead{struckmeier@fias.uni-frankfurt.de}
\begin{abstract}
The Lagrangians and Hamiltonians of classical field theory require compensating gauge fields in order to realize local internal gauge invariance with the standard transformation laws.
This principle is developed in the extended covariant canonical transformation formalism, which also accommodates changes of spacetime coordinates.
The construction yields locally U$(N)$-invariant matter--gauge-field actions and, in a separate application, diffeomorphism-covariant actions for ordinary vector fields and an independent affine connection.
The latter application adopts a covariant-derivative coupling prescription whose antisymmetric vector-momentum sector contains a direct interaction with torsion.
This interaction is not required by diffeomorphism covariance alone and does not apply to the internal gauge connection with its standard inhomogeneous transformation law.
The compensating terms are distinguished from additional invariant interactions and from the gravitational dynamics, which must be specified independently.
\end{abstract}
\section{\label{sec:intro}Introduction}
The principle of \emph{local gauge invariance} is an eminently fruitful device for constructing the interactions of the standard model.
Conventional gauge theories are commonly derived on the basis of Lagrangians of relativistic field theory~\cite{ryder06,griffiths08}.
An equivalent Hamiltonian formulation can make systematic use of the \emph{canonical transformation} formalism.
With the transformation rules for fields and their canonical conjugates derived from \emph{generating functions},
the action principle is preserved up to a boundary contribution whose variation vanishes under the stated boundary conditions.
Canonicality alone does not imply a symmetry of a prescribed Hamiltonian.
A symmetry additionally requires the transformed Hamiltonian to have the same functional form in the transformed variables.

The canonical construction identifies compensating interactions that implement this requirement for a specified transformation law.
It does not exclude further invariant interactions and does not determine all kinetic terms or coupling constants.
The internal gauge-covariant derivative and the affine covariant derivative address different transformation laws.
They must therefore be distinguished, independently of whether the affine connection has torsion.

Prior to working out local U$(N)$ gauge transformations in \sref{sec:gen-gauge-ext}, we introduce extended Lagrangians and Hamiltonians in sections~\ref{sec:gen-lagr} and~\ref{sec:gen-ham}.
These sections consider the parametrized representation of a given conventional field theory.
The functions $x^\mu(y)$ describe coordinate maps rather than independent physical changes of the metric.
The construction allows changes of parametrization without supplying gravitational field equations.
In particular, a regular coordinate transformation cannot turn a flat metric into a curved one.
The extended canonical transformation formalism is then developed in \sref{sec:gen-ct}.
Where a Legendre map is singular, the corresponding constraints must be retained or a reduced variational formulation must be used.

The U$(N)$ construction in \sref{sec:gen-gauge-ext} starts from a generating function describing the required transformation of the scalar fields in internal space.
The compensating vector fields have an inhomogeneous local transformation law.
An amended generating function then includes this transformation and determines the homogeneous transformation of their momenta.
After combining the derivative terms in the action, one obtains the first-order Yang--Mills action.
The reduced antisymmetric momentum sector is imposed explicitly rather than inferred from an unspecified Hamiltonian.
SU$(N)$ is obtained by additionally requiring the gauge matrices and their infinitesimal generators to satisfy the corresponding determinant and trace conditions.

In \sref{sec:diffeo-ext}, we consider a second application involving \emph{ordinary spacetime covectors} without the inhomogeneous internal transformation law of \sref{sec:gen-gauge-ext}.
We initially retain their unrestricted momenta and adopt an affine covariant-derivative coupling prescription.
The affine connection is treated as an independent field whose transformation is included in a generating function for a prescribed chart transition.
No metric-compatibility condition is assumed.
When a metric is present, the general setting is therefore metric-affine, with Riemann--Cartan geometry obtained only after imposing or deriving metric compatibility~\cite{hehl95}.

The final action couples the connection momentum to the curvature tensor.
For antisymmetric vector momentum, the adopted prescription also gives a direct vector--torsion interaction.
Its coefficient is not fixed by diffeomorphism invariance because that interaction is already a scalar density.
This freedom is related to the distinction between exterior differentiation and the replacement of all partial derivatives by affine covariant derivatives~\cite{delhom20}.
The gravitational dynamics and any additional invariant interactions remain separate physical inputs~\cite{struckmeier17}.
Thus the construction supplies a framework for covariant coupling rather than a derivation of a unique gravitational theory or a curvature-induced vector mass.
\section{\label{sec:gen-lagr}Extended Lagrangians $\tilde{\LC}_{\mathrm{e}}$ in the realm of classical field theory}
\subsection{Variational principle, extended set of Euler-Lagrange field equations}
The Lagrangian description of the dynamics of a continuous system (see, e.g., \cite{saletan13}) is based on the Lagrangian
density function $\tilde{\LC}$ that is supposed to convey the complete information on the given physical system.
In a first-order field theory, the Lagrangian density $\tilde{\LC}$ is defined to depend on $I=1,\ldots,N$ --- possibly interacting ---
fields $\phi^{I}(\bx)$, on the vector of independent spacetime variables $\bx\equiv x^\mu$, and on the first derivatives of the fields
$\phi^{I}$ with respect to the independent variables, i.e., on the covariant vectors ($1$-forms)
\begin{displaymath}
\pfrac{\phi^{I}}{\bx}\equiv\left(\pfrac{\phi^{I}}{ct},
\pfrac{\phi^{I}}{\pmb{x}}\right)
\equiv\left(\pfrac{\phi^{I}}{x^{\mu}}\right).
\end{displaymath}
The Euler-Lagrange field equations are then obtained as the zero of the variation $\delta S$ of the action integral
\begin{equation}\label{action-int}
S=\int_{V}\tilde{\LC}\left(\phi^{I},\pfrac{\phi^{I}}{\bx},\bx\right)\,\rmd^{4}x
\end{equation}
as
\begin{equation}\label{elgl}
\pfrac{}{x^{j}}\pfrac{\tilde{\LC}}{\left(\pfrac{\phi^{I}}{x^{j}}\right)}-\pfrac{\tilde{\LC}}{\phi^{I}}=0.
\end{equation}
In analogy to the extended formalism of point mechanics (\cite{struckmeier05,struckmeier09}), we can directly cast the
action integral from Eq.~(\ref{action-int}) into a more general form by \emph{decoupling} its integration measure from
a possibly explicit $\bx$-dependence of the Lagrangian density $\tilde{\LC}$
\begin{equation}\label{action-int1}
S=\int_{V^{\prime}}\tilde{\LC}\,\detfrac{x}{y}\,\rmd^{4}y=\int_{V^{\prime}}\tilde{\LC}^\prime\,\rmd^{4}y,\qquad\rmd^{4}x=\detfrac{x}{y}\rmd^4y.
\end{equation}
Herein, $|\partial x/\partial y|\neq0$ stands for the determinant of the Jacobi matrix that is associated with a regular transformation $\bx\mapsto\by$ of the independent variables
\begin{equation}
\detfrac{x}{y}=\fdet{x}{y}=
\left|\begin{array}{ccc}
\pfrac{x^{0}}{y^{0}}&\ldots&\pfrac{x^{0}}{y^{3}}\\
\vdots&\ddots&\vdots\\
\pfrac{x^{3}}{y^{0}}&\ldots&\pfrac{x^{3}}{y^{3}}\end{array}\right|\neq0.
\label{jacdet}
\end{equation}
We work with orientation-preserving chart transitions, so the determinant is positive.
For orientation-reversing maps the corresponding absolute Jacobian is used in the density transformation.
The quantities $\partial x^\mu/\partial y^\nu$ will be called \emph{coordinate Jacobian coefficients}.
They describe a change of coordinates rather than a change of the intrinsic spacetime geometry.

Regarding the mapping $\tilde{\LC}\mapsto\tilde{\LC}^\prime$ in Eq.~(\ref{action-int1}), we observe that
$\tilde{\LC}$ transforms under a change of the volume form as a \emph{relative scalar} of weight $w=1$, which is commonly referred to as a \emph{scalar density}.
All scalars, vectors, and tensors that transform correspondingly will be marked by a tilde in the following.

With the new volume form $\rmd^4y$, the integrand of Eq.~(\ref{action-int1}) can be thought of as defining the
\emph{extended} Lagrangian density $\tilde{\LC}_{\mathrm{e}}$, which constitutes a \emph{relative scalar} of weight $w=1$,
\begin{equation}\label{L1-def}
\tilde{\LC}_{\mathrm{e}}\left(\phi^{I}(\by),\pfrac{\phi^{I}(\by)}{\by},\bx(\by),\pfrac{\bx(\by)}{\by}\right)
=\tilde{\LC}\left(\phi^{I}(\bx),\pfrac{\phi^{I}(\bx)}{x^{\alpha}},\bx\right)\detfrac{x}{y}.
\end{equation}
With regard to the argument list of $\tilde{\LC}_{\mathrm{e}}$, the now \emph{dependent} variables
$x^{0},\ldots,x^{3}$ can be regarded as an \emph{extension} of the set of fields $\phi^{I}, I=1,\ldots,N$.
In other words, the $x^{\mu}(\by)$, defined as arbitrary functions of $\by$, are treated on equal footing with the fields $\phi^{I}(\by)$.
In terms of the extended Lagrangian $\tilde{\LC}_{\mathrm{e}}$, the action integral over $d^{4}y$ from Eq.~(\ref{action-int1})
is converted into an integral over an \emph{autonomous} Lagrangian, hence over a Lagrangian that does not \emph{explicitly} depend on its independent variables $y^{\nu}$,
\begin{equation}\label{action-int2}
S=\int_{V^{\prime}}\tilde{\LC}_{\mathrm{e}}\left(\phi^{I}(\by),
\pfrac{\phi^{I}}{\by},\bx(\by),\pfrac{\bx}{\by}\right)\,\rmd^{4}y.
\end{equation}
As this action integral has exactly the form of the initial one from Eq.~(\ref{action-int}), the Euler-Lagrange
field equations emerging from the variation of Eq.~(\ref{action-int2}) take on form of Eq.~(\ref{elgl})
\begin{equation}\label{elgl-ext1}
\pfrac{}{y^{\alpha}}\pfrac{\tilde{\LC}_{\mathrm{e}}}{\left(\pfrac{\phi^{I}}{y^{\alpha}}\right)}-\pfrac{\tilde{\LC}_{\mathrm{e}}}{\phi^{I}}=0,\qquad
\pfrac{}{y^{\alpha}}\pfrac{\tilde{\LC}_{\mathrm{e}}}{\left(\pfrac{x^{\mu}}{y^{\alpha}}\right)}-\pfrac{\tilde{\LC}_{\mathrm{e}}}{x^{\mu}}=0.
\end{equation}
With the $\phi^{I}$ embodying scalar fields, the derivatives $\partial\phi^{I}/\partial x^{\mu}$ define a covariant vector for each field $I=1,\ldots,N$.
Provided that the complete Lagrangian $\tilde{\LC}$ transforms as a scalar density, including the transformation of any background tensors,
the form of the Euler--Lagrange equations~(\ref{elgl-ext1}) is maintained under changes of coordinates.
Lorentz covariance alone does not imply this stronger property for arbitrary coordinate expressions.

If the Lagrangians $\tilde{\LC},\tilde{\LC}_{\mathrm{e}}$ are to describe the dynamics of a \emph{vector field} $a^{\mu}$
in place of a set of scalar fields $\phi^{I}$, then the Euler-Lagrange equations take on the form
\begin{equation}\label{elgl-ext1-vec}
\pfrac{}{x^{\alpha}}\pfrac{\tilde{\LC}}{\left(\pfrac{a^{\mu}}{x^{\alpha}}\right)}-\pfrac{\tilde{\LC}}{a^{\mu}}=0,\qquad
\pfrac{}{y^{\alpha}}\pfrac{\tilde{\LC}_{\mathrm{e}}}{\left(\pfrac{a^{\mu}}{y^{\alpha}}\right)}-\pfrac{\tilde{\LC}_{\mathrm{e}}}{a^{\mu}}=0.
\end{equation}
The derivatives $\partial a^{\mu}/\partial x^{\nu}$ do not separately transform as tensors.
Nevertheless, the Euler--Lagrange equations are covariant whenever the complete action is a scalar under coordinate changes.
For a vector kinetic term, this can be achieved with an exterior derivative or with an affine covariant derivative, depending on the model.
A Lorentz-covariant seed action in a reference chart does not by itself specify which extension is intended.
In a general torsionful geometry, a coordinate choice cannot set all connection coefficients to zero at an event.
The appropriate covariant actions will be constructed in the following sections.
\subsection{Equations for the extended Lagrangian $\tilde{\LC}_{\mathrm{e}}$}
In order to show that the conventional Lagrangian $\tilde{\LC}$ description of a dynamical system is compatible with the
corresponding description in terms of extended Lagrangians, we must make use of the following identities
\begin{eqnarray}
\pfrac{\phi^{I}}{x^{\mu}}=\pfrac{\phi^{I}}{y^{\alpha}}\pfrac{y^{\alpha}}{x^{\mu}}
\quad&\Longrightarrow\quad
\pfrac{\left(\pfrac{\phi^{I}}{x^{\mu}}\right)}
{\left(\pfrac{\phi^{I}}{y^{\nu}}\right)}=\delta_{\alpha}^{\nu}
\pfrac{y^{\alpha}}{x^{\mu}}=\pfrac{y^{\nu}}{x^{\mu}}\nonumber\\
\quad&\Longrightarrow\quad
\pfrac{\left(\pfrac{\phi^{I}}{x^{\alpha}}\right)}
{\left(\pfrac{y^{\nu}}{x^{\mu}}\right)}=
\pfrac{\phi^{I}}{y^{\beta}}\delta_{\nu}^{\beta}\delta_{\alpha}^{\mu}=
\pfrac{\phi^{I}}{y^{\nu}}\delta_{\alpha}^{\mu}\nonumber\\
\pfrac{x^{\nu}}{y^{\beta}}\pfrac{y^{\beta}}{x^{\mu}}=\delta_{\mu}^{\nu}
\quad&\Longrightarrow\quad
\pfrac{\left(\pfrac{y^{\nu}}{x^{\mu}}\right)}
{\left(\pfrac{x^{\alpha}}{y^{\beta}}\right)}=-\pfrac{y^{\nu}}{x^{\alpha}}
\pfrac{y^{\beta}}{x^{\mu}}\nonumber\\
\quad&\Longrightarrow\quad
\pfrac{}{y^{\alpha}}{\left(\pfrac{y^{\alpha}}{x^{\mu}}\right)}=
-\ppfrac{x^{\xi}}{y^{\alpha}}{y^{\beta}}\pfrac{y^{\alpha}}{x^{\xi}}\pfrac{y^{\beta}}{x^{\mu}}\nonumber\\
\pfrac{\left(\pfrac{\phi^{I}}{x^{\alpha}}\right)}{\left(\pfrac{x^{\mu}}{y^{\nu}}\right)}
=\pfrac{\left(\pfrac{\phi^{I}}{x^{\alpha}}\right)}{\left(\pfrac{y^{\xi}}{x^{\beta}}\right)}
\pfrac{\left(\pfrac{y^{\xi}}{x^{\beta}}\right)}{\left(\pfrac{x^{\mu}}{y^{\nu}}\right)}
&=-\pfrac{\phi^{I}}{y^{\xi}}\pfrac{y^{\xi}}{x^{\mu}}\pfrac{y^{\nu}}{x^{\alpha}}
=-\pfrac{\phi^{I}}{x^{\mu}}\pfrac{y^{\nu}}{x^{\alpha}}\nonumber\\
\pfrac{x^{\nu}}{y^{\alpha}}\pfrac{\detfrac{x}{y}}
{\left(\pfrac{x^{\mu}}{y^{\alpha}}\right)}=\delta_{\mu}^{\nu}\detfrac{x}{y}
\quad&\Longleftrightarrow\quad
\pfrac{\detfrac{x}{y}}{\left(\pfrac{x^{\mu}}{y^{\nu}}\right)}=
\pfrac{y^{\nu}}{x^{\mu}}\detfrac{x}{y}\nonumber\\
\pfrac{\detfrac{x}{y}}{y^{\mu}}=\ppfrac{x^{\alpha}}{y^{\beta}}{y^{\mu}}\pfrac{y^{\beta}}{x^{\alpha}}\,\detfrac{x}{y}
\quad&\Longrightarrow\quad
\pfrac{}{y^{\alpha}}\left(\pfrac{y^{\alpha}}{x^{\mu}}\detfrac{x}{y}\right)\equiv0,
\label{L1-identity}
\end{eqnarray}
with the last two lines following directly from the definition of the determinant.

The correlation~(\ref{L1-def}) of the extended Lagrangian $\tilde{\LC}_{\mathrm{e}}$ and conventional Lagrangian
$\tilde{\LC}$ emerges from the requirement of Eq.~(\ref{action-int1}) to yield the identical action $S$, hence to describe the same physical system.
As $\detfrac{x}{y}$ only depends on the coordinate Jacobian coefficients $\partial x^\mu/\partial y^\nu$,
the derivatives of $\tilde{\LC}_{\mathrm{e}}$ from Eq.~(\ref{L1-def}) with respect to its arguments are then
\begin{equation}\label{L1-deri0}
\pfrac{\tilde{\LC}_{\mathrm{e}}}{\phi^{I}}=\pfrac{\tilde{\LC}}{\phi^{I}}\detfrac{x}{y},\qquad
\pfrac{\tilde{\LC}_{\mathrm{e}}}{x^{\mu}}={\left.\pfrac{\tilde{\LC}}{x^{\mu}}
\right\vert}_{\mathrm{expl}}\detfrac{x}{y},
\end{equation}
where the notation ``expl'' indicates the \emph{explicit} dependence of the conventional Lagrangian $\tilde{\LC}$ on the $x^{\mu}$, and
\begin{eqnarray}
\pfrac{\tilde{\LC}_{\mathrm{e}}}{\left(\pfrac{\phi^{I}}{y^{\nu}}\right)}&=
\pfrac{\tilde{\LC}}{\left(\pfrac{\phi^{I}}{x^{\alpha}}\right)}
\pfrac{y^{\nu}}{x^{\alpha}}\detfrac{x}{y}\label{L1-deri}\\
\pfrac{\tilde{\LC}_{\mathrm{e}}}{\left(\pfrac{x^{\mu}}{y^{\nu}}\right)}&=
\tilde{\LC}\,\pfrac{\detfrac{x}{y}}{\left(\pfrac{x^{\mu}}{y^{\nu}}\right)}+
\pfrac{\tilde{\LC}}{\left(\pfrac{\phi^{I}}{x^{\alpha}}\right)}
\pfrac{\left(\pfrac{\phi^{I}}{x^{\alpha}}\right)}
{\left(\pfrac{x^{\mu}}{y^{\nu}}\right)}\detfrac{x}{y}\nonumber\\
&=\tilde{\LC}\,\pfrac{y^{\nu}}{x^{\mu}}\detfrac{x}{y}
-\pfrac{\tilde{\LC}}{\left(\pfrac{\phi^{I}}{x^{\alpha}}\right)}
\pfrac{\phi^{I}}{x^{\mu}}\pfrac{y^{\nu}}{x^{\alpha}}\detfrac{x}{y}\nonumber\\
&=\left(\delta_{\mu}^{\alpha}\tilde{\LC}-\pfrac{\tilde{\LC}}{\left(\pfrac{\phi^{I}}{x^{\alpha}}\right)}
\pfrac{\phi^{I}}{x^{\mu}}\right)\pfrac{y^{\nu}}{x^{\alpha}}\detfrac{x}{y}\nonumber\\
&=-\tilde{t}\indices{_{\mu}^{\alpha}}(\bx)\,\pfrac{y^{\nu}}{x^{\alpha}}\detfrac{x}{y}
=-\tilde{t}\indices{_{\beta}^{\nu}}(\by)\,\pfrac{y^{\beta}}{x^{\mu}},\nonumber
\end{eqnarray}
with $\tilde{t}\indices{_{\mu}^{\nu}}$ denoting the canonical energy-momentum tensor,
\begin{equation}\label{e-m-def}
\tilde{t}\indices{_{\mu}^{\nu}}(\bx)=\pfrac{\tilde{\LC}}{\left(\pfrac{\phi^{I}}{x^{\nu}}\right)}\pfrac{\phi^{I}}{x^{\mu}}-\delta_{\mu}^{\nu}\tilde{\LC}.
\end{equation}
Furthermore,
\begin{eqnarray}
\pfrac{\tilde{\LC}_{\mathrm{e}}}{\left(\pfrac{x^{\mu}}{y^{\alpha}}\right)}
\pfrac{x^{\nu}}{y^{\alpha}}&=\delta_{\mu}^{\nu}\tilde{\LC}\detfrac{x}{y}-
\pfrac{\tilde{\LC}}{\left(\pfrac{\phi^{I}}{x^{\nu}}\right)}
\pfrac{\phi^{I}}{x^{\mu}}\detfrac{x}{y}=-\tilde{t}\indices{_{\mu}^{\nu}}(\bx)\detfrac{x}{y}\nonumber\\
\pfrac{\tilde{\LC}_{\mathrm{e}}}{\left(\pfrac{x^{\alpha}}{y^{\nu}}\right)}
\pfrac{x^{\alpha}}{y^{\mu}}&=\delta_{\mu}^{\nu}\tilde{\LC}_{\mathrm{e}}-
\pfrac{\tilde{\LC}}{\left(\pfrac{\phi^{I}}{x^{\alpha}}\right)}
\pfrac{\phi^{I}}{y^{\mu}}\pfrac{y^{\nu}}{x^{\alpha}}\detfrac{x}{y}.
\label{L1-deri2}
\end{eqnarray}
The Lagrangian $\tilde{\LC}_{\mathrm{e}}$ is reproduced by summing its derivatives from Eqs.~(\ref{L1-deri}) and (\ref{L1-deri2})
\begin{equation}\label{L1-homo}
\pfrac{\tilde{\LC}_{\mathrm{e}}}{\left(\pfrac{\phi^{I}}{y^{\nu}}\right)}
\pfrac{\phi^{I}}{y^{\mu}}+\pfrac{\tilde{\LC}_{\mathrm{e}}}{\left(\pfrac{x^{\alpha}}{y^{\nu}}\right)}
\pfrac{x^{\alpha}}{y^{\mu}}\equiv\delta_{\mu}^{\nu}\tilde{\LC}_{\mathrm{e}}.
\end{equation}
Thus, if we are given a conventional Lagrangian $\tilde{\LC}$ and set up the extended Lagrangian
$\tilde{\LC}_{\mathrm{e}}=\tilde{\LC}\detfrac{x}{y}$ by multiplying $\tilde{\LC}$ with $\detfrac{x}{y}$,
then the correlation is readily shown to induce the \emph{identity}~(\ref{L1-homo}) that holds for extended Lagrangians of the parametrized form~(\ref{L1-def}).

In the derivation of Eq.~(\ref{L1-deri2}), we have made use of the fact that $\tilde{\LC}$ is a \emph{conventional} Lagrangian,
i.e.\ a Lagrangian that depends on the coordinate Jacobian coefficients $\partial x^\mu/\partial y^\nu$ only \emph{indirectly}
via the reparametrization condition~(\ref{L1-def}) applied to its velocities,
\begin{displaymath}
\pfrac{\phi^{I}}{x^{\mu}}=\pfrac{\phi^{I}}{y^{\alpha}}\pfrac{y^{\alpha}}{x^{\mu}}.
\end{displaymath}
Defining the extended energy-momentum tensor similarly to the conventional one from Eq.~(\ref{e-m-def})
\begin{equation}\label{e-m-def-ext}
\LT\indices{_{\mu}^{\nu}}=\pfrac{\tilde{\LC}_{\mathrm{e}}}{\left(\pfrac{\phi^{I}}{y^{\nu}}\right)}
\pfrac{\phi^{I}}{y^{\mu}}+\pfrac{\tilde{\LC}_{\mathrm{e}}}{\left(\pfrac{x^{\alpha}}{y^{\nu}}\right)}
\pfrac{x^{\alpha}}{y^{\mu}}-\delta_{\mu}^{\nu}\tilde{\LC}_{\mathrm{e}},
\end{equation}
we thus have
\begin{displaymath}
\LT\indices{_{\mu}^{\nu}}\equiv0.
\end{displaymath}
For the extended Lagrangian $\tilde{\LC}_{\mathrm{e}}$, all elements of the extended energy-momentum tensor~(\ref{e-m-def-ext}) thus always vanish.

For a given conventional Lagrangian $\tilde{\LC}$, the related extended Lagrangian $\tilde{\LC}_{\mathrm{e}}=\tilde{\LC}\detfrac{x}{y}$
is a homogeneous function of degree $4$ in the simultaneous derivative variables, as for any $k>0$ the mapping
\begin{eqnarray*}
\pfrac{\phi^{I}}{y^{\nu}}&\mapsto& k\,\pfrac{\phi^{I}}{y^{\nu}},\quad\pfrac{x^{\mu}}{y^{\nu}}\mapsto k\,\pfrac{x^{\mu}}{y^{\nu}}\\
\pfrac{\phi^I}{x^\mu}=\pfrac{\phi^I}{y^\alpha}\pfrac{y^\alpha}{x^\mu}
&\mapsto& k\pfrac{\phi^I}{y^\alpha}\,k^{-1}\pfrac{y^\alpha}{x^\mu}
=\pfrac{\phi^I}{x^\mu}.
\end{eqnarray*}
yields
\begin{eqnarray*}
\tilde{\LC}_{\mathrm{e}}\left(\phi^{I},k\,\pfrac{\phi^{I}}{y^{\nu}},x^{\mu},k\,\pfrac{x^{\mu}}{y^{\nu}}\right)&=
\tilde{\LC}\left(\phi^{I},\pfrac{\phi^{I}}{x^{\mu}},x^{\mu}\right)\left|k\pfrac{\bx}{\by}\right|\\
&=\tilde{\LC}\left(\phi^{I},\pfrac{\phi^{I}}{x^{\mu}},x^{\mu}\right)k^{4}\detfrac{x}{y}\\
&=k^{4}\tilde{\LC}_{\mathrm{e}}\left(\phi^{I},\pfrac{\phi^{I}}{y^{\nu}},x^{\mu},\pfrac{x^{\mu}}{y^{\nu}}\right).
\end{eqnarray*}
Taking the trace of Eq.~(\ref{L1-homo}) gives the Euler identity for this degree-four homogeneity.
The untraced equation is the stronger matrix identity associated with the parametrized density.
Both follow from the construction of $\tilde{\LC}_{\mathrm{e}}$.
Then, the left-hand side Euler-Lagrange equation from Eq.~(\ref{elgl-ext1}) is equivalent to the conventional Euler-Lagrange equation~(\ref{elgl})
\begin{eqnarray*}\fl
\qquad\qquad\!\!\pfrac{\tilde{\LC}_{\mathrm{e}}}{\phi^{I}}=
\pfrac{}{y^{j}}\pfrac{\tilde{\LC}_{\mathrm{e}}}{\left(\pfrac{\phi^{I}}{y^{j}}\right)}&=
\pfrac{}{y^{j}}\left(\pfrac{\tilde{\LC}}{\left(\pfrac{\phi^{I}}{x^{i}}\right)}\right)
\pfrac{y^{j}}{x^{i}}\detfrac{x}{y}+\pfrac{\tilde{\LC}}{\left(\pfrac{\phi^{I}}{x^{i}}\right)}
\underbrace{\pfrac{}{y^{j}}\left(\pfrac{y^{j}}{x^{i}}\detfrac{x}{y}\right)}_{\equiv0}\\
&=\pfrac{}{x^{i}}\left(\pfrac{\tilde{\LC}}{\left(\pfrac{\phi^{I}}{x^{i}}\right)}\right)
\detfrac{x}{y}
\stackrel{\mathrm{Eq.}~(\ref{L1-deri0})}{=}\pfrac{\tilde{\LC}}{\phi^{I}}\detfrac{x}{y}\\
\Longrightarrow\qquad\pfrac{\tilde{\LC}}{\phi^{I}}
&=\pfrac{}{x^{i}}\left(\pfrac{\tilde{\LC}}{\left(\pfrac{\phi^{I}}{x^{i}}\right)}\right).
\end{eqnarray*}
The right-hand side Euler-Lagrange equation from Eq.~(\ref{elgl-ext1}) does not provide any information on the dynamics of the coordinate Jacobian coefficients.
Yet, this equation quantifies the divergence of the $\mu$-th column of the energy-momentum tensor due to an explicit dependence of $\tilde{\LC}$ on $x^{\mu}$,
\begin{eqnarray}
\pfrac{\tilde{\LC}_{\mathrm{e}}}{x^{\mu}}=\pfrac{}{y^{\alpha}}\pfrac{\tilde{\LC}_{\mathrm{e}}}{\left(\pfrac{x^{\mu}}{y^{\alpha}}\right)}&=
-\pfrac{\tilde{t}\indices{_{\mu}^{\beta}}(\bx)}{y^{\alpha}}\pfrac{y^{\alpha}}{x^{\beta}}\detfrac{x}{y}-\tilde{t}\indices{_{\mu}^{\beta}}(\bx)
\underbrace{\pfrac{}{y^{\alpha}}\left(\pfrac{y^{\alpha}}{x^{\beta}}\detfrac{x}{y}\right)}_{\equiv0}\nonumber\\
&=-\pfrac{\tilde{t}\indices{_{\mu}^{\beta}}}{x^{\beta}}\detfrac{x}{y}
\stackrel{\mathrm{Eq.}~(\ref{L1-deri0})}{=}{\left.\pfrac{\tilde{\LC}}{x^{\mu}}
\right\vert}_{\mathrm{expl}}\detfrac{x}{y}\nonumber\\
\Longrightarrow\qquad{\left.\pfrac{\tilde{\LC}}{x^{\mu}}\right\vert}_{\mathrm{expl}}
&=-\pfrac{\tilde{t}\indices{_{\mu}^{\beta}}}{x^{\beta}}.\label{L-conv-expl}
\end{eqnarray}
The explicit coordinate dependence gives the source term in the canonical energy-momentum balance law.
Its interpretation as an external force depends on the specified background and coordinate representation.
\section{\label{sec:gen-ham}Extended Hamiltonians $\tilde{\HC}_{\mathrm{e}}$ in classical field theory}
\subsection{Extended canonical field equations}
For a covariant Hamiltonian description, the canonical momenta are tensor densities of weight one in their respective charts,
\begin{equation}\label{p1-def}
\tilde\pi_I^\mu(x)
=\pfrac{\tilde{\LC}}{\left(\pfrac{\phi^I}{x^\mu}\right)},
\qquad
\tilde\pi_I^\nu(y)
=\pfrac{\tilde{\LC}_{\mathrm e}}{\left(\pfrac{\phi^I}{y^\nu}\right)}.
\end{equation}
Equation~(\ref{L1-deri}) gives
\begin{equation}\label{p1-def1}
\tilde\pi_I^\nu(y)
=\tilde\pi_I^\alpha(x)\pfrac{y^\nu}{x^\alpha}\detfrac{x}{y}.
\end{equation}
When a metric is present, the associated absolute vector is $\pi_I^\mu=\tilde\pi_I^\mu/\sqrt{-g}$.
The tilde is retained on canonical momenta and Hamiltonians in all density-valued equations below.

Similar to the momentum field $\tilde{\pi}_{I}^{\nu}(\by)$ constituting the dual counterpart of the Lagrangian variable
$\partial\phi^{I}(\by)/\partial y^{\nu}$, we define the canonical variable $\tilde{t}\indices{_{\mu}^{\nu}}$ as the dual quantity to $\partial x^{\mu}(\by)/\partial y^{\nu}$.
It follows --- similarly to Eq.~(\ref{p1-def}) --- from the partial derivative of the extended Lagrangian density
$\tilde{\LC}_{\mathrm{e}}$ with respect to $\partial x^{\mu}/\partial y^{\nu}$,
\begin{equation}\label{pext-def}
\tilde{t}\indices{_{\mu}^{\nu}}=-\pfrac{\tilde{\LC}_{\mathrm{e}}}{\left(
\pfrac{x^{\mu}}{y^{\nu}}\right)}=
\tilde{t}\indices{_{\alpha}^{\nu}}(\by)\,\pfrac{y^{\alpha}}{x^{\mu}}=
\tilde{t}\indices{_{\mu}^{\alpha}}(\bx)\,\pfrac{y^{\nu}}{x^{\alpha}}\detfrac{x}{y}.
\end{equation}
Note that in this form, the indices of $\tilde{t}\indices{_{\mu}^{\nu}}$ on the left-hand side refer to different coordinate frames.
We can now introduce both the De~Donder--Weyl Hamiltonian~\cite{dedonder30,weyl35} $\tilde{\HC}$ and the extended Hamiltonian
$\tilde{\HC}_{\mathrm{e}}$ as the covariant Legendre transforms of the Lagrangian $\tilde{\LC}$ and of the extended Lagrangian $\tilde{\LC}_{\mathrm{e}}$, respectively
\begin{eqnarray}
\tilde{\HC}\big(\phi^{I},\tilde{\pi}_{I}^{\nu},x^{\mu}\big)&=\tilde{\pi}_{I}^{\alpha}\pfrac{\phi^{I}}{x^{\alpha}}-
\tilde{\LC}\left(\phi^{I},\pfrac{\phi^{I}}{x^{\mu}},x^{\mu}\right)\label{H-def}\\
\tilde{\HC}_{\mathrm{e}}\big(\phi^{I},\tilde{\pi}_{I}^{\nu},x^{\mu},
\tilde{t}\indices{_{\mu}^{\nu}}\big)&=
\tilde{\pi}_{I}^{\alpha}\pfrac{\phi^{I}}{y^{\alpha}}-
\tilde{t}\indices{_{\beta}^{\alpha}}\pfrac{x^{\beta}}{y^{\alpha}}-
\tilde{\LC}_{\mathrm{e}}\left(\phi^{I},\pfrac{\phi^{I}}{y^{\nu}},x^{\mu},
\pfrac{x^{\mu}}{y^{\nu}}\right).\label{H1-def}
\end{eqnarray}
The derivatives on the right-hand sides of Eqs.~(\ref{H-def}) and~(\ref{H1-def}) must be eliminated in favor of the momenta wherever the corresponding Legendre map is locally invertible.
Until that elimination has been performed, these equations are identities on the Legendre image rather than derivative-free phase-space functions.
For a singular map, the Hamiltonian formulation includes the primary constraints and their multipliers instead of using an unrestricted inverse.
Tracing Eq.~(\ref{L1-homo}) gives, on the Legendre image,
\begin{displaymath}
\tilde{\pi}_I^\mu\pfrac{\phi^I}{y^\mu}
-\tilde{t}\indices{_\alpha^\mu}\pfrac{x^\alpha}{y^\mu}
=4\tilde{\LC}_{\mathrm e},
\qquad
\tilde{\HC}_{\mathrm e}=3\tilde{\LC}_{\mathrm e}.
\end{displaymath}
In particular, the parametrized field theory does not imply $\tilde{\HC}_{\mathrm e}=0$.

From the correlation~(\ref{L1-def}) of the extended Lagrangian density $\tilde{\LC}_{\mathrm{e}}$
to the conventional Lagrangian $\tilde{\LC}$, we obtain the correlation of extended and conventional Hamiltonians as
\begin{eqnarray*}
\tilde{\HC}_{\mathrm{e}}&=\tilde{\pi}_{I}^{\alpha}(\by)\pfrac{\phi^{I}}{y^{\alpha}}-
\tilde{t}\indices{_{\beta}^{\alpha}}\pfrac{x^{\beta}}{y^{\alpha}}-\tilde{\LC}\detfrac{x}{y}\\
&=\cancel{\tilde{\pi}_{I}^{\beta}(\bx)\pfrac{y^{\alpha}}{x^{\beta}}\pfrac{\phi^{I}}{y^{\alpha}}\detfrac{x}{y}}
-\tilde{t}\indices{_{\beta}^{\alpha}}\pfrac{x^{\beta}}{y^{\alpha}}
-\cancel{\tilde{\pi}_{I}^{\alpha}(\bx)\pfrac{\phi^{I}}{x^{\alpha}}\detfrac{x}{y}}+\tilde{\HC}\detfrac{x}{y}.
\end{eqnarray*}
As the first and the third term on the right-hand side cancel, the extended Hamiltonian density
$\tilde{\HC}_{\mathrm{e}}\big|_y$ is related to the conventional Hamiltonian $\tilde{\HC}\big|_x$ by
\begin{equation}\label{H1-def1}
\tilde{\HC}_{\mathrm{e}}\big(\phi^{I},\tilde{\bpi}_{I}(y),\bx,\tilde{\bt}_{\mu}\big)
=\tilde{\HC}\big(\phi^{I},\tilde{\bpi}_{I}(x),\bx\big)\detfrac{x}{y}
-\tilde{t}\indices{_{\beta}^{\alpha}}\pfrac{x^{\beta}}{y^{\alpha}}.
\end{equation}
The momenta on the right-hand side of Eq.~(\ref{H1-def1}) must be expressed through Eq.~(\ref{p1-def1}), and all remaining derivatives are understood on the same Legendre branch.
The displayed correlation is not a prescription to hold its Jacobians fixed in phase-space differentiation.
According to Eq.~(\ref{L1-deri}), the function $\partial\tilde{\LC}_{\mathrm{e}}/\partial(\partial x^{\mu}/\partial y^{\nu})$
is related to the component $\tilde{t}\indices{_{\mu}^{\nu}}$ of the energy-momentum tensor.
Thus, $\tilde{t}\indices{_{\mu}^{\nu}}$ as the dual counterpart of $\partial x^{\mu}/\partial y^{\nu}$ is
\begin{equation}\label{pext-def1}
\tilde{t}\indices{_{\mu}^{\nu}}(\by)=\tilde{t}\indices{_{\alpha}^{\nu}}\,
\pfrac{x^{\alpha}}{y^{\mu}}\qquad\Longleftrightarrow\qquad
\tilde{t}\indices{_{\mu}^{\nu}}(\bx)=\tilde{t}\indices{_{\mu}^{\alpha}}\,
\pfrac{x^{\nu}}{y^{\alpha}}\,\detfrac{y}{x}.
\end{equation}
Expressed in terms of the trace density $\tilde t\indices{_{\alpha}^{\alpha}}(\bx)$, the extended Hamiltonian density $\tilde{\HC}_{\mathrm{e}}$ from Eq.~(\ref{H1-def1}) is given by
\begin{equation}\label{H1-def2}
\tilde{\HC}_{\mathrm{e}}\Big|_y=\left(\tilde{\HC}\Big|_x-\tilde{t}\indices{_{\alpha}^{\alpha}}(\bx)\right)\detfrac{x}{y}.
\end{equation}
According to Eq.~(\ref{p1-def}) and the conventional set of Euler-Lagrange equations~(\ref{elgl}),
the conventional Hamiltonian $\tilde{\HC}$, defined in Eq.~(\ref{H-def}), satisfies the conventional set of covariant canonical equations
\begin{equation}\label{fgln}
\pfrac{\tilde{\HC}}{\tilde{\pi}_{I}^{\mu}}=\pfrac{\phi^{I}}{x^{\mu}},\qquad
\pfrac{\tilde{\HC}}{\phi^{I}}=-\pfrac{\tilde{\LC}}{\phi^{I}}=
-\pfrac{\tilde{\pi}_{I}^{\alpha}}{x^{\alpha}},\qquad
{\left.\pfrac{\tilde{\HC}}{x^{\mu}}\right\vert}_{\mathrm{expl}}=
-{\left.\pfrac{\tilde{\LC}}{x^{\mu}}\right\vert}_{\mathrm{expl}}.
\end{equation}
On a regular Legendre branch, the differential of Eq.~(\ref{H1-def}) gives the canonical equations without treating the coordinate Jacobian coefficients as independent fixed parameters.
The terms proportional to variations of the eliminated derivatives cancel by the definitions of the momenta.
One obtains
\begin{eqnarray}
\pfrac{\tilde{\HC}_{\mathrm e}}{\tilde{\pi}_I^\nu}
&=\pfrac{\phi^I}{y^\nu},
\qquad
\pfrac{\tilde{\HC}_{\mathrm e}}{\tilde{t}\indices{_\mu^\nu}}
=-\pfrac{x^\mu}{y^\nu},\nonumber\\
\pfrac{\tilde{\HC}_{\mathrm e}}{\phi^I}
&=-\pfrac{\tilde{\pi}_I^\nu}{y^\nu},
\qquad
\pfrac{\tilde{\HC}_{\mathrm e}}{x^\mu}
=\pfrac{\tilde{t}\indices{_\mu^\nu}}{y^\nu}.
\label{ext-fgln}
\end{eqnarray}
These equations are supplemented by the relevant constraint equations in a singular formulation.
The momentum-divergence equations alone do not determine every component of a momentum field.
Nevertheless, a divergence-free addition to a momentum generally changes the companion field equation and is not an arbitrary canonical gauge freedom.
Likewise, an improvement of a Noether current must not be identified with an unrestricted shift of the canonical energy-momentum variable.

The embedding equation is
\begin{equation}\label{caneq-tr-H1}
\pfrac{x^\mu}{y^\nu}
=-\pfrac{\tilde{\HC}_{\mathrm e}}{\tilde{t}\indices{_\mu^\nu}}.
\end{equation}
It relates the embedding derivatives to the Legendre variables.
Its interpretation as a canonical equation requires the velocity elimination just described.
Merely differentiating the explicit $-\tilde t\,\partial x/\partial y$ term in Eq.~(\ref{H1-def1}) while holding all Jacobians fixed would instead produce a bookkeeping identity.
The parametrized theory is equivalent to the original theory and supplies no independent gravitational field equation.

The conjugate field equation quantifies the divergence of the energy-momentum tensor density
$\tilde{t}\indices{_{\mu}^{\nu}}$ from the $x^{\mu}$-dependent terms of the extended Hamiltonian~(\ref{H1-def1})
\begin{equation}\label{caneq-tr-H2}
\pfrac{\tilde{t}\indices{_{\mu}^{\alpha}}}{y^{\alpha}}=
\pfrac{\tilde{\HC}_{\mathrm{e}}}{x^{\mu}}=
{\left.\pfrac{\tilde{\HC}}{x^{\mu}}\right\vert}_{\mathrm{expl}}\detfrac{x}{y}.
\end{equation}
On the other hand, we have
\begin{eqnarray*}
\pfrac{\tilde{t}\indices{_{\mu}^{\alpha}}}{y^{\alpha}}&=
\pfrac{}{y^{\alpha}}\left(\tilde{t}\indices{_{\mu}^{\beta}}
\pfrac{y^{\alpha}}{x^{\beta}}\detfrac{x}{y}\right)=
\pfrac{\tilde{t}\indices{_{\mu}^{\beta}}}{y^{\alpha}}
\pfrac{y^{\alpha}}{x^{\beta}}\detfrac{x}{y}+
\tilde{t}\indices{_{\mu}^{\beta}}\underbrace{\pfrac{}{y^{\alpha}}\left(
\pfrac{y^{\alpha}}{x^{\beta}}\detfrac{x}{y}\right)}_{=0}\\
&=\pfrac{\tilde{t}\indices{_{\mu}^{\alpha}}}{x^{\alpha}}\detfrac{x}{y},
\end{eqnarray*}
so that
\begin{displaymath}
\pfrac{\tilde{t}\indices{_{\mu}^{\alpha}}}{x^{\alpha}}=
{\left.\pfrac{\tilde{\HC}}{x^{\mu}}\right\vert}_{\mathrm{expl}},
\end{displaymath}
in agreement with Eqs.~(\ref{L-conv-expl}) and (\ref{fgln}).
The extended Hamiltonian~(\ref{H1-def1}) thus simply reproduces the field equations of the conventional covariant Hamiltonian
$\tilde{\HC}$ while allowing arbitrary regular changes of parametrization.
This property permits extended canonical transformations of the embeddings in addition to transformations of the fields and their conjugates.
It does not identify the embeddings with dynamical metric variables.

The action integral from Eq.~(\ref{action-int2}) can be equivalently expressed in terms of the extended Hamiltonian density
$\tilde{\HC}_{\mathrm{e}}$ by applying the Legendre transform~(\ref{H1-def})
\begin{equation}\label{action-int3}
S=\int_{V^{\prime}}\left(
\tilde{\pi}_{I}^{\beta}\pfrac{\phi^{I}}{y^{\beta}}-
\tilde{t}\indices{_{\alpha}^{\beta}}\pfrac{x^{\alpha}}{y^{\beta}}-
\tilde{\HC}_{\mathrm{e}}\left(\phi^{I},\tilde{\pi}_{I}^{\nu},x^{\mu},%
\tilde{t}\indices{_{\mu}^{\nu}}\right)\right)\rmd^{4}y.
\end{equation}
This representation of the action integral forms the basis on which extended canonical transformations will be defined in~\sref{sec:gen-ct}.

In case that the extended Lagrangian describes the dynamics of a (covariant) vector field, $a_{\mu}$,
rather than the dynamics of a set of scalar fields, $\phi^{I}$, the canonical momentum fields are to be defined as
\begin{equation}\label{p1-def-vec}
\tilde{p}^{\mu\nu}(\by)=
\pfrac{\tilde{\LC}_{\mathrm{e}}}{\left(\pfrac{a_{\mu}}{y^{\nu}}\right)}.
\end{equation}
From the transformation rule for a covariant vector field
\begin{displaymath}
a_{\mu}(\bx)=a_{\alpha}(\by)\pfrac{y^{\alpha}}{x^{\mu}},
\end{displaymath}
the rule for its partial derivatives follows as
\begin{displaymath}
\pfrac{a_{\mu}(\bx)}{x^{\nu}}=\pfrac{a_{\alpha}(\by)}{y^{\beta}}
\pfrac{y^{\alpha}}{x^{\mu}}\pfrac{y^{\beta}}{x^{\nu}}+
a_{\alpha}(\by)\ppfrac{y^{\alpha}}{x^{\mu}}{x^{\nu}}.
\end{displaymath}
Thus
\begin{displaymath}
\pfrac{\left(\pfrac{a_{\mu}(\bx)}{x^{\nu}}\right)}
{\left(\pfrac{a_{\xi}(\by)}{y^{\eta}}\right)}=
\delta_{\alpha}^{\xi}\delta_{\beta}^{\eta}\,
\pfrac{y^{\alpha}}{x^{\mu}}\pfrac{y^{\beta}}{x^{\nu}}=
\pfrac{y^{\xi}}{x^{\mu}}\pfrac{y^{\eta}}{x^{\nu}}.
\end{displaymath}
The canonical momentum fields $\tilde{p}^{\mu\nu}(\by)$ then transform according to
\begin{eqnarray*}
\tilde{p}^{\mu\nu}(\by)&=\pfrac{\tilde{\LC}\detfrac{x}{y}}
{\left(\pfrac{a_{\alpha}(\bx)}{x^{\beta}}\right)}
\pfrac{\left(\pfrac{a_{\alpha}(\bx)}{x^{\beta}}\right)}
{\left(\pfrac{a_{\mu}(\by)}{y^{\nu}}\right)}\\
&=\tilde p^{\alpha\beta}(\bx)\,\pfrac{y^{\mu}}{x^{\alpha}}
\pfrac{y^{\nu}}{x^{\beta}}\detfrac{x}{y}.
\end{eqnarray*}
The momentum densities in the two charts are related in the same way as Eq.~(\ref{p1-def1}), with an additional contravariant transformation for the vector-field index.
An associated absolute tensor can be defined by $p^{\mu\nu}=\tilde p^{\mu\nu}/\sqrt{-g}$ when a metric is present.
\subsection{Extended energy-momentum tensor}
The Hamiltonian formulation of the extended version of the energy-momentum tensor from Eqs.~(\ref{e-m-def-ext}) is defined by
\begin{equation}\label{e-m-def-h-ext}
\LT\indices{_{\mu}^{\nu}}=\tilde{\HC}_{\mathrm{e}}\,\delta_{\mu}^{\nu}+\tilde{\pi}_{I}^{\nu}
\pfrac{\phi^{I}}{y^{\mu}}-\delta_{\mu}^{\nu}\,\tilde{\pi}_{I}^{\alpha}\pfrac{\phi^{I}}{y^{\alpha}}-
\tilde{t}\indices{_{\alpha}^{\nu}}\pfrac{x^{\alpha}}{y^{\mu}}+
\delta_{\mu}^{\nu}\,\tilde{t}\indices{_{\alpha}^{\beta}}\pfrac{x^{\alpha}}{y^{\beta}}.
\end{equation}
Inserting into~(\ref{e-m-def-h-ext}) the definitions~(\ref{p1-def}) and (\ref{pext-def}) of the canonical conjugates
$\tilde{\pi}_{I}^{\nu}$ and $\tilde{t}_{\mu}^{\nu}$ of the variables $\phi^{I}$ and $x^{\mu}$, respectively,
the identity~(\ref{L1-homo}) that holds for the extended Lagrangian $\tilde{\LC}_{\mathrm{e}}$ is expressed in terms of
$\LT\indices{_{\mu}^{\nu}}$ with $\tilde{\HC}_{\mathrm{e}}$ the extended Hamiltonian from Eq.~(\ref{H1-def2}) as
\begin{displaymath}
\LT\indices{_{\mu}^{\nu}}\equiv0.
\end{displaymath}
For an extended Hamiltonian obtained from Eq.~(\ref{L1-def}), all elements of the extended energy-momentum
expression~(\ref{e-m-def-h-ext}) vanish on the Legendre image, which is in accordance with the corresponding Lagrangian formulation from Eq.~(\ref{e-m-def-ext}).

Making use of the extended set of canonical field equations~(\ref{ext-fgln}), we can directly verify that the
divergence of the extended energy-momentum tensor $\LT\indices{_{\mu}^{\nu}}(\by)$ vanishes for all indexes~$\mu$,
as expected for the \emph{identity} $\LT\indices{_{\mu}^{\nu}}\equiv0$
\begin{eqnarray*}
\pfrac{\LT\indices{_{\mu}^{\alpha}}(\by)}{y^{\alpha}}&=
\left(\pfrac{\tilde{\HC}_{\mathrm{e}}}{\phi^{I}}\pfrac{\phi^{I}}{y^{\alpha}}+
\pfrac{\tilde{\HC}_{\mathrm{e}}}{\tilde{\pi}_{I}^{\beta}}\pfrac{\tilde{\pi}_{I}^{\beta}}{y^{\alpha}}+
\pfrac{\tilde{\HC}_{\mathrm{e}}}{x^{\beta}}\pfrac{x^{\beta}}{y^{\alpha}}+
\pfrac{\tilde{\HC}_{\mathrm{e}}}{\tilde{t}\indices{_{\xi}^{\beta}}}
\pfrac{\tilde{t}\indices{_{\xi}^{\beta}}}{y^{\alpha}}\right)\delta_{\mu}^{\alpha}\\
&\qquad\mbox{}+\pfrac{\tilde{\pi}_{I}^{\alpha}}{y^{\alpha}}\pfrac{\phi^{I}}{y^{\mu}}+
\cancel{\tilde{\pi}_{I}^{\alpha}\,\ppfrac{\phi^{I}}{y^{\mu}}{y^{\alpha}}}-
\delta_{\mu}^{\alpha}\pfrac{\tilde{\pi}_{I}^{\beta}}{y^{\alpha}}\pfrac{\phi^{I}}{y^{\beta}}-
\cancel{\delta_{\mu}^{\beta}\tilde{\pi}_{I}^{\alpha}\,\ppfrac{\phi^{I}}{y^{\alpha}}{y^{\beta}}}\\
&\qquad\mbox{}-\pfrac{\tilde{t}\indices{_{\alpha}^{\beta}}}{y^{\beta}}\pfrac{x^{\alpha}}{y^{\mu}}-
\cancel{\tilde{t}\indices{_{\alpha}^{\beta}}\,\ppfrac{x^{\alpha}}{y^{\mu}}{y^{\beta}}}+
\delta_{\mu}^{\beta}\pfrac{\tilde{t}\indices{_{\alpha}^{\xi}}}{y^{\beta}}\pfrac{x^{\alpha}}{y^{\xi}}+
\cancel{\delta_{\mu}^{\beta}\tilde{t}\indices{_{\alpha}^{\xi}}\,\ppfrac{x^{\alpha}}{y^{\xi}}{y^{\beta}}}\\
&=-\pfrac{\tilde{\pi}_{I}^{\xi}}{y^{\xi}}\pfrac{\phi^{I}}{y^{\mu}}+
\pfrac{\phi^{I}}{y^{\alpha}}\pfrac{\tilde{\pi}_{I}^{\alpha}}{y^{\mu}}+
\pfrac{\tilde{t}\indices{_{\alpha}^{\xi}}}{y^{\xi}}\pfrac{x^{\alpha}}{y^{\mu}}-
\pfrac{x^{\alpha}}{y^{\xi}}\pfrac{\tilde{t}\indices{_{\alpha}^{\xi}}}{y^{\mu}}\\
&\qquad\mbox{}+\pfrac{\tilde{\pi}_{I}^{\beta}}{y^{\beta}}\pfrac{\phi^{I}}{y^{\mu}}-
\pfrac{\tilde{\pi}_{I}^{\xi}}{y^{\mu}}\pfrac{\phi^{I}}{y^{\xi}}-
\pfrac{\tilde{t}\indices{_{\alpha}^{\beta}}}{y^{\beta}}\pfrac{x^{\alpha}}{y^{\mu}}+
\pfrac{\tilde{t}\indices{_{\alpha}^{\xi}}}{y^{\mu}}\pfrac{x^{\alpha}}{y^{\xi}}\\
&\equiv0.
\end{eqnarray*}
The vanishing of this extended expression is a parametrization identity.
It does not imply an additional physical energy conservation law for the unparametrized subsystem.
The subsystem balance law remains Eq.~(\ref{L-conv-expl}), including any explicit dependence on the original coordinates.
\section{\label{sec:gen-ct}Extended
canonical transformations}
\subsection{Generating function of type $\bFC_{1}\big|_y$}
We may set up the condition for canonical transformations that include a mapping $\bx(\by)\mapsto\bX(\by)$
of the parametrizations of source and target systems, $\bx$ and $\bX$, respectively, with $y^{\mu}$ being the common independent variables of both systems.
We again require the action integral --- this time in the formulation of Eqs.~(\ref{action-int2}) and
(\ref{action-int3}) --- to be equivalent under the transformation, up to a boundary contribution
\begin{eqnarray*}
\delta S&=\delta\int_{V^{\prime}}\left(\tilde{\pi}_{I}^{\beta}\pfrac{\phi^{I}}{y^{\beta}}-
\tilde{t}\indices{_{\alpha}^{\beta}}\pfrac{x^{\alpha}}{y^{\beta}}-
\tilde{\HC}_{\mathrm{e}}\left(\phi^{I},\tilde{\pi}_{I}^{\nu},x^{\mu},%
\tilde{t}\indices{_{\mu}^{\nu}}\right)\right)\rmd^{4}y\\
&=\delta\int_{V^{\prime}}\left(
\tilde{\Pi}_{I}^{\beta}\pfrac{\Phi^{I}}{y^{\beta}}-
\tilde{T}\indices{_{\alpha}^{\beta}}\pfrac{X^{\alpha}}{y^{\beta}}-
\tilde{\HC}_{\mathrm{e}}^\prime\left(\Phi^{I},\tilde{\Pi}_{I}^{\nu},X^{\mu},%
\tilde{T}\indices{_{\mu}^{\nu}}\right)\right)\rmd^{4}y.
\end{eqnarray*}
That means, its integrand is determined up to the divergence of a $4$-vector density
$\FC^\mu_{1}\big|_y=\FC_{1}^\mu(\phi^{I},\Phi^{I},x^{\mu},X^{\mu})$ of the sets of original fields $\phi^{I}(\by)$
and transformed fields $\Phi^{I}(\by)$ in conjunction with the original variables $x^{\mu}(\by)$ and transformed variables, $X(\by)$
\begin{eqnarray}
\tilde{\LC}_{\mathrm{e}}&=\tilde{\LC}_{\mathrm{e}}^{\prime}+\pfrac{\FC_{1}^{\alpha}}{y^{\alpha}}
\qquad\Longleftrightarrow\nonumber\\
\tilde{\pi}_{I}^{\beta}&\pfrac{\phi^{I}}{y^{\beta}}-
\tilde{t}\indices{_{\alpha}^{\beta}}\pfrac{x^{\alpha}}{y^{\beta}}-
\tilde{\HC}_{\mathrm{e}}\left(\phi^{I},\tilde{\pi}_{I}^{\nu},%
x^{\mu},\tilde{t}\indices{_{\mu}^{\nu}}\right)\nonumber\\
&=\tilde{\Pi}_{I}^{\beta}\pfrac{\Phi^{I}}{y^{\beta}}-
\tilde{T}\indices{_{\alpha}^{\beta}}\pfrac{X^{\alpha}}{y^{\beta}}-
\tilde{\HC}_{\mathrm{e}}^{\prime}\left(\Phi^{I},\tilde{\Pi}_{I}^{\nu},%
X^{\mu},\tilde{T}\indices{_{\mu}^{\nu}}\right)+
\pfrac{\FC_{1}^{\beta}}{y^{\beta}}.\label{intbed-ext}
\end{eqnarray}
The independent variables $y^{\mu}$ are not transformed.
Boundary variations are assumed to vanish, or fields and variations are taken with compact support.
The generating function in this subsection has no explicit $y$ dependence.
An additional explicit dependence would give $\tilde{\HC}_{\mathrm e}'=\tilde{\HC}_{\mathrm e}+\Hexpl{\FC_1^\mu}{y^\mu}$.
The transformation must be locally invertible and must preserve any constraints of the chosen Legendre formulation.
Moreover, the ordinary divergence of a vector density $\FC_{1}^\mu(\phi^{I},\Phi^{I},x^{\mu},X^{\mu})$ constitutes a scalar density
\begin{equation}\label{divF-ext}
\pfrac{\FC_{1}^{\beta}}{y^{\beta}}=\pfrac{\FC_{1}^{\beta}}{\phi^{I}}\pfrac{\phi^{I}}{y^{\beta}}+
\pfrac{\FC_{1}^{\beta}}{\Phi^{I}}\pfrac{\Phi^{I}}{y^{\beta}}+
\pfrac{\FC_{1}^{\beta}}{x^{\alpha}}\pfrac{x^{\alpha}}{y^{\beta}}+
\pfrac{\FC_{1}^{\beta}}{X^{\alpha}}\pfrac{X^{\alpha}}{y^{\beta}}.
\end{equation}
Comparing the coefficients of Eqs.~(\ref{intbed-ext}) and (\ref{divF-ext}), we find the \emph{extended}
local coordinate representation of the field transformation rules induced by the extended generating function
$\FC_{1}^\mu\big|_y=\FC_{1}^\mu(\phi^{I},\Phi^{I},x^{\mu},X^{\mu})$
\begin{eqnarray*}
\tilde{\pi}_{I}^{\mu}=\pfrac{\FC_{1}^{\mu}}{\phi^{I}},\quad
\tilde{\Pi}_{I}^{\mu}=-\pfrac{\FC_{1}^{\mu}}{\Phi^{I}},\quad
\tilde{t}\indices{_{\nu}^{\mu}}=-\pfrac{\FC_{1}^{\mu}}{x^{\nu}},\quad
\tilde{T}\indices{_{\nu}^{\mu}}=\pfrac{\FC_{1}^{\mu}}{X^{\nu}},\quad
\tilde{\HC}_{\mathrm{e}}^{\prime}=\tilde{\HC}_{\mathrm{e}},
\end{eqnarray*}
hence
\begin{eqnarray}
\tilde{\pi}_{I}^{\mu}(\bx)&=\hphantom{-}\pfrac{\FC_{1}^{\alpha}}{\phi^{I}}\pfrac{x^\mu}{y^\alpha}\detfrac{y}{x},\qquad\thickspace
\tilde{\Pi}_{I}^{\mu}(\bX)&=-\pfrac{\FC_{1}^{\alpha}}{\Phi^{I}}\pfrac{X^\mu}{y^\alpha}\detfrac{y}{X},\nonumber\\
\tilde{t}\indices{_{\nu}^{\mu}}(\bx)&=-\pfrac{\FC_{1}^{\alpha}}{x^{\nu}}\pfrac{x^{\mu}}{y^{\alpha}}\detfrac{y}{x},\qquad
\tilde{T}\indices{_{\nu}^{\mu}}(\bX)&=\hphantom{-}\pfrac{\FC_{1}^{\alpha}}{X^{\nu}}\pfrac{X^{\mu}}{y^{\alpha}}\detfrac{y}{X}.
\label{genF1-ext}
\end{eqnarray}
The \emph{values} of the extended Hamiltonians $\tilde{\HC}_{\mathrm{e}}^\prime,\tilde{\HC}_{\mathrm{e}}$
are thus conserved under extended canonical transformations if they refer to the same space-time event $y$.
Hence, the new extended Hamiltonian density $\tilde{\HC}_{\mathrm{e}}^{\prime}(\Phi^{I},\tilde{\Pi}_{I},X^{\mu},\tilde{T}\indices{_{\mu}^{\nu}})$
is obtained by simply expressing the original extended Hamiltonian $\tilde{\HC}_{\mathrm{e}}(\phi^{I},\tilde{\pi}_{I}^{\nu},x^{\mu},\tilde{t}\indices{_{\mu}^{\nu}})$
in terms of the transformed fields $\Phi^{I}$, $\tilde{\Pi}_{I}^{\nu}$, the transformed space-time location $X^{\mu}$,
and the transformed energy-momentum tensor density, $\tilde{T}\indices{_{\mu}^{\nu}}$.

The transformation rule for the conventional Hamiltonian density $\tilde{\HC}\mapsto\tilde{\HC}^{\prime}$ is obtained from
$\tilde{\HC}_{\mathrm{e}}^{\prime}=\tilde{\HC}_{\mathrm{e}}$ by inserting the definition~(\ref{H1-def2}) of the extended Hamiltonian
\begin{equation}\label{ctH1}
\left(\tilde{\HC}^{\prime}-\tilde{T}\indices{_{\alpha}^{\alpha}}(X)\right)\detfrac{X}{y}=
\left(\tilde{\HC}-\tilde{t}\indices{_{\alpha}^{\alpha}}(x)\right)\detfrac{x}{y},
\end{equation}
hence
\begin{displaymath}\fl
\left(\tilde{\HC}^\prime-\tilde{T}\indices{_{\alpha}^{\alpha}}(X)\right)\detfrac{X}{x}=\tilde{\HC}-\tilde{t}\indices{_{\alpha}^{\alpha}}(x),\qquad
\tilde{\HC}^\prime-\tilde{T}\indices{_{\alpha}^{\alpha}}(X)=\left(\tilde{\HC}-\tilde{t}\indices{_{\alpha}^{\alpha}}(x)\right)\detfrac{x}{X}.
\end{displaymath}
This means for the extended Hamiltonians $\tilde{\HC}_{\mathrm{e}}^{\prime}$ and $\tilde{\HC}_{\mathrm{e}}$ that they transform as scalar densities if taken at $X$ and $x$, respectively
\begin{equation}\label{He-trans}
\tilde{\HC}_{\mathrm{e}}^{\prime}\Big|_X=\tilde{\HC}_{\mathrm{e}}\Big|_x\detfrac{x}{X}.
\end{equation}
\subsection{Transformation of the extended energy-momentum expression $\LT\indices{_{\mu}^{\nu}}$ under extended canonical transformations}
Inserting the transformation rules for the original variables pertaining to the generating function of type
$\bFC_{1}$ from Eq.~(\ref{genF1-ext}) into the Hamiltonian representation of the extended energy-momentum tensor~(\ref{e-m-def-h-ext}), one finds
\begin{displaymath}
\LT\indices{_{\mu}^{\nu}}=\tilde{\HC}_{\mathrm{e}}\,\delta_{\mu}^{\nu}+
\pfrac{\FC_{1}^{\nu}}{\phi^{I}}\pfrac{\phi^{I}}{y^{\mu}}+
\pfrac{\FC_{1}^{\nu}}{x^{\alpha}}\pfrac{x^{\alpha}}{y^{\mu}}-
\delta_{\mu}^{\nu}\left(\pfrac{\FC_{1}^{\alpha}}{\phi^{I}}\pfrac{\phi^{I}}{y^{\alpha}}+
\pfrac{\FC_{1}^{\alpha}}{x^{\beta}}\pfrac{x^{\beta}}{y^{\alpha}}\right).
\end{displaymath}
The derivatives of $\FC_{1}^\nu$ with respect to the original variables $\phi^{I}$ and $x^{\alpha}$ can be converted
into derivatives with respect to the transformed variables $\Phi^{I}$ and $X^{\alpha}$ according to Eq.~(\ref{divF-ext}),
which yields with $\tilde{\HC}_{\mathrm{e}}=\tilde{\HC}_{\mathrm{e}}^{\prime}$
\begin{displaymath}\fl
\quad\LT\indices{_{\mu}^{\nu}}=\tilde{\HC}_{\mathrm{e}}^{\prime}\,\delta_{\mu}^{\nu}+
\pfrac{\FC_{1}^{\nu}}{y^{\mu}}-
\pfrac{\FC_{1}^{\nu}}{\Phi^{I}}\pfrac{\Phi^{I}}{y^{\mu}}-
\pfrac{\FC_{1}^{\nu}}{X^{\alpha}}\pfrac{X^{\alpha}}{y^{\mu}}-
\delta_{\mu}^{\nu}\left(\pfrac{\FC_{1}^{\alpha}}{y^{\alpha}}-
\pfrac{\FC_{1}^{\alpha}}{\Phi^{I}}\pfrac{\Phi^{I}}{y^{\alpha}}-
\pfrac{\FC_{1}^{\alpha}}{X^{\beta}}\pfrac{X^{\beta}}{y^{\alpha}}\right)\!.
\end{displaymath}
Inserting now the transformation rules for the new variables from Eq.~(\ref{genF1-ext}), this gives
\begin{eqnarray*}\fl
\qquad\qquad\LT\indices{_{\mu}^{\nu}}&=\tilde{\HC}_{\mathrm{e}}^{\prime}\,\delta_{\mu}^{\nu}+
\pfrac{\FC_{1}^{\nu}}{y^{\mu}}+
\tilde{\Pi}_{I}^{\nu}\pfrac{\Phi^{I}}{y^{\mu}}-
\tilde{T}\indices{_{\alpha}^{\nu}}\pfrac{X^{\alpha}}{y^{\mu}}-
\delta_{\mu}^{\nu}\!\left(\pfrac{\FC_{1}^{\alpha}}{y^{\alpha}}+
\tilde{\Pi}_{I}^{\alpha}\pfrac{\Phi^{I}}{y^{\alpha}}-
\tilde{T}\indices{_{\beta}^{\alpha}}\pfrac{X^{\beta}}{y^{\alpha}}\right)\\\fl
&=(\LT\indices{_{\mu}^{\nu}})^{\prime}+\tilde{S}\indices{_{\mu}^{\nu}},\qquad
\tilde{S}\indices{_{\mu}^{\nu}}=\pfrac{\FC_{1}^{\nu}}{y^{\mu}}-
\delta_{\mu}^{\nu}\pfrac{\FC_{1}^{\alpha}}{y^{\alpha}}.\fl
\end{eqnarray*}
The \emph{value} of the extended energy-momentum tensor $\LT\indices{_{\mu}^{\nu}}$ is thus maintained
under the canonical transformation generated by $\FC_{1}^{\nu}$ up to the tensor $\tilde{S}\indices{_{\mu}^{\nu}}$.
The divergence of $\tilde{S}\indices{_{\mu}^{\nu}}$ obviously vanishes identically
\begin{displaymath}
\pfrac{\tilde{S}\indices{_{\mu}^{\alpha}}}{y^{\alpha}}=\ppfrac{\FC_{1}^{\alpha}}{y^{\mu}}{y^{\alpha}}-
\delta_{\mu}^{\beta}\ppfrac{\FC_{1}^{\alpha}}{y^{\alpha}}{y^{\beta}}\equiv0.
\end{displaymath}
The displayed relation describes a Noether-current improvement in the fixed $y$ chart.
It does not establish equality of the two canonical currents as defined, nor does it authorize an arbitrary change of a canonical momentum or of a gravitational source.
The divergence is unchanged,
\begin{displaymath}
\pfrac{\LT\indices{_\mu^\alpha}}{y^\alpha}
=\pfrac{(\LT\indices{_\mu^\alpha})'}{y^\alpha}.
\end{displaymath}
Integrated charges agree when the corresponding boundary contribution vanishes.
If both descriptions are restricted to the parametrized class with vanishing extended current, the improvement must also vanish on that class.

\subsection{Alternative formulation for generating functions of type $\bFC_1\big|_X$}
For the gauge constructions below, it is convenient to work directly with two prescribed coordinate charts rather than to vary their embeddings.
The functions $x=x(X)$ and their Jacobians are fixed transformation coefficients.
The canonical action identity then reads
\begin{eqnarray}
\left(\tilde\pi_I^\alpha\pfrac{\phi^I}{x^\alpha}-\tilde{\HC}\right)\detfrac{x}{X}
&=\tilde\Pi_I^\beta\pfrac{\Phi^I}{X^\beta}-\tilde{\HC}'+\pfrac{\FC_1^\beta}{X^\beta}.
\label{intbed-ext-2}
\end{eqnarray}
Here $\FC_1^\mu$ depends on the original and transformed fields and may depend explicitly on $X$ through the prescribed chart map.
Its divergence is
\begin{eqnarray}
\pfrac{\FC_1^\beta}{X^\beta}
&=\pfrac{\FC_1^\beta}{\phi^I}\pfrac{x^\alpha}{X^\beta}\pfrac{\phi^I}{x^\alpha}
+\pfrac{\FC_1^\beta}{\Phi^I}\pfrac{\Phi^I}{X^\beta}
+\Hexpl{\FC_1^\beta}{X^\beta}.
\label{divF-ext-2}
\end{eqnarray}
Comparison of the field-derivative coefficients gives
\begin{eqnarray}
\tilde\pi_I^\mu(x)
&=\pfrac{\FC_1^\beta}{\phi^I}\pfrac{x^\mu}{X^\beta}\detfrac{X}{x},
\qquad
\tilde\Pi_I^\mu(X)=-\pfrac{\FC_1^\mu}{\Phi^I},\nonumber\\
\tilde{\HC}'(X)
&=\tilde{\HC}(x)\detfrac{x}{X}
+\Hexpl{\FC_1^\beta}{X^\beta}.
\label{genF1-ext-2}
\end{eqnarray}
The last term contains derivatives of all prescribed chart coefficients but not derivatives of the canonical variables.
The Hamiltonian transformation law can equivalently be written as
\begin{equation}\label{ctH1-2}
\tilde{\HC}'(X)-\tilde{\HC}(x)\detfrac{x}{X}
=\Hexpl{\FC_1^\beta}{X^\beta}.
\end{equation}
This is the fixed-chart counterpart of the extended identity~(\ref{ctH1}).
It does not assign independent canonical variations to $x(X)$ or to its derivatives.

\subsection{Generating function of type $\bFC_2\big|_X$}
The generating function can alternatively be expressed in terms of the original fields and the new momenta.
Where the relevant partial Legendre transformation exists, it is defined by
\begin{equation}\label{legendre1-ext}
\FC_2^\mu(\phi^I,\tilde\Pi_I^\nu,X)
=\FC_1^\mu(\phi^I,\Phi^I,X)+\tilde\Pi_I^\mu\Phi^I.
\end{equation}
The resulting transformation rules are
\begin{eqnarray}
\tilde\pi_I^\mu(x)
&=\pfrac{\FC_2^\alpha}{\phi^I}\pfrac{x^\mu}{X^\alpha}\detfrac{X}{x},
\qquad
\Phi^I\delta_\nu^\mu=\pfrac{\FC_2^\mu}{\tilde\Pi_I^\nu},\nonumber\\
\tilde{\HC}'(X)
&=\tilde{\HC}(x)\detfrac{x}{X}
+\Hexpl{\FC_2^\mu}{X^\mu}.
\label{genF2-ext}
\end{eqnarray}
A type-two generating function can also be specified directly when the corresponding type-one representation is singular.
Not every vector-valued function is admissible.
The momentum derivatives must define the same transformed field in every component, including the off-diagonal conditions
in Eq.~(\ref{genF2-ext}), and the induced field--momentum map must be locally invertible.
The linear momentum dependence used below satisfies these conditions on the unrestricted component space.
For tensor fields, the field label in these rules includes their tensor indices.
If a momentum is subsequently restricted by symmetry, the induced transformation must preserve that restriction and the final action must be varied within the restricted space.
For example, differentiation with respect to an antisymmetric momentum gives
\begin{displaymath}
\pfrac{\tilde P^{\rho\mu}}{\tilde P^{\alpha\nu}}
=\onehalf\left(\delta^\rho_\alpha\delta^\mu_\nu
-\delta^\rho_\nu\delta^\mu_\alpha\right).
\end{displaymath}
Consequently, the unrestricted component rules must not be reused without projection after imposing antisymmetry.
\section{General local U($N$) gauge transformation in the extended canonical formalism\label{sec:gen-gauge-ext}}
\subsection{External gauge field}
We now apply the fixed-chart rules of Eq.~(\ref{genF2-ext}) to the internal transformation of complex scalar fields.
The spacetime coordinates are unchanged in this section, so $X=x$.
The conventional scalar-density Hamiltonian of the seed system is denoted by $\tilde{\HC}_0$.
The corresponding action is
\begin{equation}\label{action-int4}
S=\int_V\left(
\tilde{\bar\pi}_I^\mu\pfrac{\phi^I}{x^\mu}
+\pfrac{\bar\phi_I}{x^\mu}\tilde\pi^{I\mu}
-\tilde{\HC}_0\right)\rmd^4x.
\end{equation}
This is the unparametrized theory whose extended representation was described in the preceding sections.
On its Legendre image and in the chart $y=x$, Eq.~(\ref{H1-def2}) gives
\begin{equation}\label{trivial-H1}
\tilde{\HC}_0
=\tilde{\HC}_{\mathrm e}\big|_{y=x}
+\tilde t\indices{_\alpha^\alpha}.
\end{equation}
The fixed-chart variational principle is Eq.~(\ref{action-int4}).
Equation~(\ref{trivial-H1}) is not used to vary a velocity-containing expression as an unconstrained phase-space Hamiltonian.

We assume the seed Hamiltonian to be form-invariant under global transformations
\begin{equation}\label{pointtra-ext}
\Phi^I=u\indices{^I_K}\phi^K,
\qquad
\bar\Phi_I=\bar\phi_K\bar u\indices{^K_I}.
\end{equation}
The matrix $U$ is unitary,
\begin{displaymath}
UU^\dagger=U^\dagger U=\Eins,
\qquad
\bar{\bPhi}\bPhi=\bar{\bphi}\bphi.
\end{displaymath}
The transformation is generated by
\begin{eqnarray}
\FC_2^\mu&=\tilde{\bar{\bPi}}^{\mu}U\bphi+\bar{\bphi}\,U^\dagger\tilde{\bPi}^\mu\nonumber\\
&=\tilde{\bar{\Pi}}_K^\mu u\indices{^K_J}\phi^J
+\bar{\phi}_K\bar u\indices{^K_J}\tilde\Pi^{J\mu}.
\label{extgen-general-gt}
\end{eqnarray}
For $U=U(x)$, all spacetime dependence of $U$ is explicit dependence of the fixed-chart generator.
There is no dependence on a varied embedding derivative.
The transformation rules give
\begin{eqnarray}
\tilde{\bar\pi}_I^\mu
&=\tilde{\bar\Pi}_J^\mu u\indices{^J_I},
\qquad
\tilde\pi^{I\mu}=\bar u\indices{^I_J}\tilde\Pi^{J\mu},\nonumber\\
\phi^I&=\bar u\indices{^I_J}\Phi^J,
\qquad
\bar\phi_I=\bar\Phi_Ju\indices{^J_I},\nonumber\\
\tilde\Pi^{I\mu}&=u\indices{^I_J}\tilde\pi^{J\mu},
\qquad
\tilde{\bar\Pi}_I^\mu=\tilde{\bar\pi}_J^\mu\bar u\indices{^J_I}.
\label{pointtra-rules-ext}
\end{eqnarray}
Complex fields and their conjugates may be varied independently during the calculation, with the physical complex-conjugation conditions imposed on the resulting fields.
The canonical Hamiltonian transformation law is
\begin{eqnarray}
\tilde{\HC}'-\tilde{\HC}
&=\Hexpl{\FC_2^\alpha}{x^\alpha}\nonumber\\
&=\left(\tilde{\bar\Pi}_K^\alpha\Phi^J
-\bar\Phi_K\tilde\Pi^{J\alpha}\right)
\pfrac{u\indices{^K_I}}{x^\alpha}\bar u\indices{^I_J}.
\label{pointtra-rules-ext-1}
\end{eqnarray}
For a global transformation, the additional term vanishes,
\begin{equation}\label{pointtra-rules-ext-global}
\pfrac{u\indices{^K_I}}{x^\alpha}=0
\quad\Longrightarrow\quad
\tilde{\HC}'-\tilde{\HC}=0.
\end{equation}
The converse is not asserted for a special field configuration.
For a general local transformation, the seed Hamiltonian alone need not retain its functional form.
Canonicality still preserves the action principle with the transformed Hamiltonian prescribed by Eq.~(\ref{pointtra-rules-ext-1}).

To obtain local form-invariance, we define the amended Hamiltonians with the same functional form,
\begin{eqnarray}
\tilde{\HC}_1'
&=\tilde{\HC}_0(\bar\Phi,\tilde\Pi,\Phi,\tilde{\bar\Pi},x)
+\rmi q\left(\tilde{\bar\Pi}_K^\alpha\Phi^J
-\bar\Phi_K\tilde\Pi^{J\alpha}\right)A\indices{^K_J_\alpha},
\label{H1p-definition}\\
\tilde{\HC}_1
&=\tilde{\HC}_0(\bar\phi,\tilde\pi,\phi,\tilde{\bar\pi},x)
+\rmi q\left(\tilde{\bar\pi}_K^\alpha\phi^J
-\bar\phi_K\tilde\pi^{J\alpha}\right)a\indices{^K_J_\alpha}.
\label{H-ext-tilde-0}
\end{eqnarray}
Here $q$ is a nonzero real coupling constant.
Since the derivative-free seed terms are invariant under the algebraic unitary transformation at each point, the difference is
\begin{eqnarray}
\tilde{\HC}_1'-\tilde{\HC}_1
&=\rmi q\left(\tilde{\bar\Pi}_K^\alpha\Phi^J
-\bar\Phi_K\tilde\Pi^{J\alpha}\right)
\left(A\indices{^K_J_\alpha}
-u\indices{^K_I}a\indices{^I_L_\alpha}\bar u\indices{^L_J}\right).
\label{form-invariant-1}
\end{eqnarray}
This reproduces Eq.~(\ref{pointtra-rules-ext-1}) when
\begin{equation}\label{gauge-tra1-ext}
A\indices{^K_J_\mu}
=u\indices{^K_I}a\indices{^I_L_\mu}\bar u\indices{^L_J}
+\frac{1}{\rmi q}\pfrac{u\indices{^K_I}}{x^\mu}\bar u\indices{^I_J}.
\end{equation}
Thus form-invariance means equality of the functional forms in the transformed variables, not numerical equality of the two Hamiltonians under a local transformation.
The additional Hamiltonian term is the one required by the explicit coordinate dependence of the generator.

Unitarity implies that $(\partial_\mu U)U^\dagger$ is skew-Hermitian.
Consequently, Eq.~(\ref{gauge-tra1-ext}) preserves the chosen Hermitian character of the gauge matrices $a_\mu$.
The connection takes values in the Hermitian representation of the Lie algebra of U$(N)$.
For SU$(N)$ the matrices are additionally traceless and $\det U=1$.
At this stage the gauge field is external.
To include its dynamics, the generating function and the seed Hamiltonian must also include the gauge field and its conjugate momentum.
\subsection{Including the gauge field dynamics}
The Hermitian gauge field matrix $a\indices{^K_J_\mu}$ is now treated as a dynamical field.
We first use unrestricted momentum components in the generating-function calculation.
The first-order seed action is
\begin{eqnarray}
S=\int_V\rmd^4x\left(
\tilde{\bar\pi}_I^\beta\pfrac{\phi^I}{x^\beta}
+\pfrac{\bar\phi_I}{x^\beta}\tilde\pi^{I\beta}
+\tilde p\indices{^J_K^\alpha^\beta}\pfrac{a\indices{^K_J_\alpha}}{x^\beta}
-\tilde{\HC}_0\right).
\label{action-int5}
\end{eqnarray}
The choice of gauge-field kinetic terms in $\tilde{\HC}_0$ remains a physical input.
The transformation property of the gauge fields can be described by amending the generating function from Eq.~(\ref{extgen-general-gt}) as follows
\begin{eqnarray}
\FC_2^\mu&=\tilde{\bar{\bPi}}^{\mu}U\bphi+\bar{\bphi}\,U^\dagger\tilde{\bPi}^\mu
+\operatorname{tr}\left[\tilde{\bP}^{\alpha\mu}
\left(Ua_\alpha U^\dagger+\frac{1}{\rmi q}\pfrac{U}{x^\alpha}U^\dagger\right)\right]
\nonumber\\
&=\tilde{\bar\Pi}_K^\mu u\indices{^K_J}\phi^J
+\bar\phi_K\bar u\indices{^K_J}\tilde\Pi^{J\mu}\nonumber\\
&\quad+\tilde P\indices{^J_K^\alpha^\mu}
\left(u\indices{^K_I}a\indices{^I_L_\alpha}\bar u\indices{^L_J}
+\frac{1}{\rmi q}\pfrac{u\indices{^K_I}}{x^\alpha}\bar u\indices{^I_J}\right).
\label{extgen2-general-gt}
\end{eqnarray}
The functions $\tilde{P}\indices{^J_K^\alpha^\mu}$ denote the canonical conjugates of the transformed gauge-field components $A\indices{^K_J_\alpha}$.
The completed seed density $\tilde{\HC}_0$ is required to be invariant under the full local transformation.
In the minimal model it depends on the gauge momentum, the scalar fields and their momenta, and any background metric, but not explicitly on the gauge potential.
Any additional dependence must satisfy the same invariance requirement.
The additional transformation rules derived from the amended generating function~(\ref{extgen2-general-gt}) are
\begin{eqnarray}
\delta_{\nu}^{\mu}A\indices{^K_J_\alpha}&=
\pfrac{\FC_{2}^{\mu}}{\tilde{P}\indices{^J_K^\alpha^\nu}}=
\delta_{\nu}^{\mu}\left(u\indices{^K_I}\,a\indices{^I_L_\alpha}\,\bar{u}\indices{^L_J}+
\frac{1}{\rmi q}\pfrac{u\indices{^K_I}}{x^{\alpha}}\,\bar{u}\indices{^I_J}\right)\nonumber\\
\tilde{p}\indices{^N_M^\beta^\mu}&=\pfrac{\FC_{2}^{\mu}}{a\indices{^M_N_\beta}}=
\tilde{P}\indices{^J_K^\alpha^\mu}u\indices{^K_I}\,\delta^I_M\,\delta^L_N\,
\delta_{\alpha}^{\beta}\,\bar{u}\indices{^L_J}=\bar{u}\indices{^N_J}\,
\tilde{P}\indices{^J_K^\beta^\mu}\,u\indices{^K_M}\nonumber\\
\tilde{\HC}_2'-\tilde{\HC}_2
&=\Hexpl{\FC_2^\mu}{x^\mu}.\nonumber\\
\label{pointtra-rules2-ext}
\end{eqnarray}
The canonical transformation obviously reproduces the required transformation rule for the gauge fields from Eq.~(\ref{gauge-tra1-ext})
as the amended generating function~(\ref{extgen2-general-gt}) was \emph{deliberately} constructed accordingly.
In contrast to the gauge fields $a_{IJ\nu}$, their canonical momenta, $p_{JI}^{\nu\mu}$ transform \emph{homogeneously}.
The explicit divergence in Eq.~(\ref{pointtra-rules2-ext}) gives
\begin{eqnarray*}
&\tilde{\HC}_{2}^{\prime}-\tilde{\HC}_{2}
=\left(\tilde{\bar{\Pi}}_{K}^{\beta}\Phi^{J}-\bar{\Phi}_{K}\,
\tilde{\Pi}^{J\beta}\right)\pfrac{u\indices{^K_I}}{x^{\beta}}\,\bar{u}\indices{^I_J}\\
&\!\!\!\!\!\!\!\!\!\!\!\!\!\mbox{}+\tilde{P}\indices{^J_K^\alpha^\beta}
\left(\pfrac{u\indices{^K_I}}{x^{\beta}}\,a\indices{^I_L_\alpha}\,\bar{u}\indices{^L_J}+
u\indices{^K_I}a\indices{^I_L_\alpha}\,\pfrac{\bar{u}\indices{^L_J}}{x^{\beta}}+
\frac{1}{\rmi q}\pfrac{u\indices{^K_I}}{x^{\alpha}}
\,\pfrac{\bar{u}\indices{^I_J}}{x^{\beta}}+\frac{1}{\rmi q}
\ppfrac{u\indices{^K_I}}{x^{\alpha}}{x^{\beta}}\,\bar{u}\indices{^I_J}\right).
\end{eqnarray*}
The second derivative term is \emph{symmetric} in the indices $\alpha$ and $\beta$.
We therefore split $\tilde{P}\indices{^J_K^\alpha^\beta}$ into a symmetric $\tilde{P}\indices{^J_K^{(\alpha\beta)}}$
and a skew-symmetric $\tilde{P}\indices{^J_K^{[\alpha\beta]}}$ part in $\alpha$ and $\beta$
\begin{equation}\label{skew-symmetric-tilde-p}
\tilde{P}\indices{^J_K^\alpha^\beta}=
\tilde{P}\indices{^J_K^{(\alpha\beta)}}+
\tilde{P}\indices{^J_K^{[\alpha\beta]}},
\end{equation}
where
\begin{displaymath}
\tilde{P}\indices{^J_K^{[\alpha\beta]}}=\onehalf\left(
\tilde{P}\indices{^J_K^\alpha^\beta}-\tilde{P}\indices{^J_K^\beta^\alpha}\right),\qquad
\tilde{P}\indices{^J_K^{(\alpha\beta)}}=\onehalf\left(
\tilde{P}\indices{^J_K^\alpha^\beta}+\tilde{P}\indices{^J_K^\beta^\alpha}\right).
\end{displaymath}
The product of the second partial derivative term with $\tilde{P}\indices{^J_K^{[\alpha\beta]}}$ vanishes
\begin{displaymath}
\tilde{P}\indices{^J_K^{[\alpha\beta]}}\,\ppfrac{u\indices{^K_I}}{x^{\alpha}}{x^{\beta}}\equiv0.
\end{displaymath}
The transformation rule for the Hamiltonians then splits into three groups of terms
\begin{eqnarray*}\fl
\qquad&\tilde{\HC}_{2}^{\prime}-\tilde{\HC}_{2}
=\left(\tilde{\bar{\Pi}}_{K}^{\beta}\Phi^{J}-\bar{\Phi}_{K}\,
\tilde{\Pi}^{J\beta}\right)\pfrac{u\indices{^K_I}}{x^{\beta}}\,\bar{u}\indices{^I_J}\\\fl
&\quad\mbox{}+\tilde{P}\indices{^J_K^{[\alpha\beta]}}\left(
\pfrac{u\indices{^K_I}}{x^{\beta}}\,a\indices{^I_L_\alpha}
\,\bar{u}\indices{^L_J}+u\indices{^K_I}a\indices{^I_L_\alpha}
\,\pfrac{\bar{u}\indices{^L_J}}{x^{\beta}}+
\frac{1}{\rmi q}\pfrac{u\indices{^K_I}}{x^{\alpha}}
\,\pfrac{\bar{u}\indices{^I_J}}{x^{\beta}}\right)\\\fl
&\quad\mbox{}+\tilde{P}\indices{^J_K^{(\alpha\beta)}}\left(
\pfrac{u\indices{^K_I}}{x^{\beta}}\,a\indices{^I_L_\alpha}
\,\bar{u}\indices{^L_J}+u\indices{^K_I}a\indices{^I_L_\alpha}
\,\pfrac{\bar{u}\indices{^L_J}}{x^{\beta}}+
\frac{1}{\rmi q}\pfrac{u\indices{^K_I}}{x^{\alpha}}
\,\pfrac{\bar{u}\indices{^I_J}}{x^{\beta}}+\frac{1}{\rmi q}
\ppfrac{u\indices{^K_I}}{x^{\alpha}}{x^{\beta}}\,\bar{u}\indices{^I_J}\right).
\end{eqnarray*}
The $u\indices{^K_I}$-dependence of the terms involving the fields $\Phi^{I}$ and their conjugates, $\tilde{\bar{\Pi}}_{I}^{\mu}$,
can be replaced according to Eq.~(\ref{form-invariant-1}) by a gauge field dependence
\begin{eqnarray*}
&\quad\left(\tilde{\bar{\Pi}}_{K}^{\beta}\Phi^{J}-
\bar{\Phi}_{K}\,\tilde{\Pi}^{J\beta}\right)
\pfrac{u\indices{^K_I}}{x^{\beta}}\,\bar{u}\indices{^I_J}\\
&=\left(\tilde{\bar{\Pi}}_{K}^{\beta}\Phi^{J}-
\bar{\Phi}_{K}\,\tilde{\Pi}^{J\beta}\right)\rmi q\,A\indices{^K_J_\beta}-
\left(\tilde{\bar{\pi}}_{K}^{\beta}\phi^{J}-
\bar{\phi}_{K}\,\tilde{\pi}^{J\beta}\right)\rmi q\,a\indices{^K_J_\beta}.
\end{eqnarray*}
In the second group, the $u\indices{^K_I}$-dependence of the terms proportional to $\tilde{P}\indices{^J_K^{[\beta\alpha]}}$
can be eliminated by means of the transformation rule~(\ref{gauge-tra1-ext}) for the gauge fields, which reads, equivalently
\begin{eqnarray*}
\pfrac{u\indices{^K_I}}{x^{\mu}}&=\rmi q\left(A\indices{^K_J_\mu}u\indices{^J_I}-u\indices{^K_J}\,a\indices{^J_I_\mu}\right)\\
\pfrac{\bar{u}\indices{^I_J}}{x^{\mu}}&=\rmi q\left(a\indices{^I_K_\mu}\,\bar{u}\indices{^K_J}-\bar{u}\indices{^I_K}A\indices{^K_J_\mu}\right).
\end{eqnarray*}
In conjunction with the transformation rule for the canonical momenta $\tilde{p}\indices{^J_K^\nu^\mu}$ from Eqs.~(\ref{pointtra-rules2-ext}), one finds
\begin{eqnarray*}\fl
&\quad\tilde{P}\indices{^J_K^{[\alpha\beta]}}\left(\pfrac{u\indices{^K_I}}{x^{\beta}}\,
a\indices{^I_L_\alpha}\,\bar{u}\indices{^L_J}+u\indices{^K_L}a\indices{^L_I_\alpha}\,
\pfrac{\bar{u}\indices{^I_J}}{x^{\beta}}+\frac{1}{\rmi q}
\pfrac{u\indices{^K_I}}{x^{\alpha}}\,\pfrac{\bar{u}\indices{^I_J}}{x^{\beta}}\right)\\\fl
&=\rmi q\,\tilde{P}\indices{^J_K^{[\alpha\beta]}}\left(-A\indices{^K_I_\alpha}\,A\indices{^I_J_\beta}+
u\indices{^K_N}\,a\indices{^N_I_\alpha}\,a\indices{^I_L_\beta}\,\bar{u}\indices{^L_J}\right)\\[\smallskipamount]\fl
&=\rmi q\,\tilde{p}\indices{^J_K^{[\alpha\beta]}}\,a\indices{^K_I_\alpha}\,a\indices{^I_J_\beta}-
\rmi q\,\tilde{P}\indices{^J_K^{[\alpha\beta]}}\,A\indices{^K_I_\alpha}\,A\indices{^I_J_\beta}\\[\smallskipamount]\fl
&=\onehalf\rmi q\,\tilde{p}\indices{^J_K^\alpha^\beta}\left(
a\indices{^K_I_\alpha}\,a\indices{^I_J_\beta}-a\indices{^K_I_\beta}\,a\indices{^I_J_\alpha}\right)-
\onehalf\rmi q\,\tilde{P}\indices{^J_K^\alpha^\beta}\left(
A\indices{^K_I_\alpha}\,A\indices{^I_J_\beta}-A\indices{^K_I_\beta}\,A\indices{^I_J_\alpha}\right).
\end{eqnarray*}
Finally, the $u\indices{^K_I}$-dependence of the terms proportional to $\tilde{P}\indices{^J_K^{(\alpha\beta)}}$
can be eliminated as well by means of both the canonical transformation rules for the gauge fields~(\ref{gauge-tra1-ext})
and that for the momenta from equations~(\ref{pointtra-rules2-ext})
\begin{eqnarray*}\fl
&\tilde{P}\indices{^J_K^{(\alpha\beta)}}\left(
\pfrac{u\indices{^K_I}}{x^{\beta}}\,a\indices{^I_L_\alpha}
\,\bar{u}\indices{^L_J}+u\indices{^K_I}a\indices{^I_L_\alpha}
\,\pfrac{\bar{u}\indices{^L_J}}{x^{\beta}}+
\frac{1}{\rmi q}\pfrac{u\indices{^K_I}}{x^{\alpha}}
\,\pfrac{\bar{u}\indices{^I_J}}{x^{\beta}}+\frac{1}{\rmi q}
\ppfrac{u\indices{^K_I}}{x^{\alpha}}{x^{\beta}}\,\bar{u}\indices{^I_J}\right)\\\fl
&=\tilde{P}\indices{^J_K^{(\alpha\beta)}}\left(
\pfrac{A\indices{^K_J_\alpha}}{x^{\beta}}-u\indices{^K_I}\,
\pfrac{a\indices{^I_L_\alpha}}{x^{\beta}}\,\bar{u}\indices{^L_J}\right)\\\fl
&=\onehalf\tilde{P}\indices{^J_K^\alpha^\beta}\left(
\pfrac{A\indices{^K_J_\alpha}}{x^{\beta}}+\pfrac{A\indices{^K_J_\beta}}{x^{\alpha}}\right)-
\onehalf\tilde{p}\indices{^J_K^\alpha^\beta}\left(\pfrac{a\indices{^K_J_\alpha}}{x^{\beta}}+
\pfrac{a\indices{^K_J_\beta}}{x^{\alpha}}\right).
\end{eqnarray*}
The symmetric sum of the gauge-field derivatives does not transform homogeneously under the local internal transformation.
Its contribution compensates the corresponding nonhomogeneous terms in the unrestricted action identity.
Once combined with the canonical kinetic term, it leaves the antisymmetric field strength.

In all three groups, the transformation coefficients have been eliminated in favor of the original and transformed field expressions.
These expressions have the same functional form and occur with opposite signs.
This identifies the required compensating contribution, without fixing additional invariant terms.
Summarizing, the transformation rule for the Hamiltonians is given by
\begin{eqnarray*}\fl
\tilde{\HC}_{2}^{\prime}&-\tilde{\HC}_{2}
=\rmi q\left(\tilde{\bar{\Pi}}_{K}^{\beta}\Phi^{J}-
\bar{\Phi}_{K}\,\tilde{\Pi}^{J\beta}\right)A\indices{^K_J_\beta}-
\rmi q\left(\tilde{\bar{\pi}}_{K}^{\beta}\phi^{J}-
\bar{\phi}_{K}\,\tilde{\pi}^{J\beta}\right)a\indices{^K_J_\beta}\\\fl
&-\onehalf\rmi q\,\tilde{P}\indices{^J_K^\alpha^\beta}\left(
A\indices{^K_I_\alpha}\,A\indices{^I_J_\beta}-A\indices{^K_I_\beta}\,A\indices{^I_J_\alpha}\right)+
\onehalf\rmi q\,\tilde{p}\indices{^J_K^\alpha^\beta}\left(
a\indices{^K_I_\alpha}\,a\indices{^I_J_\beta}-a\indices{^K_I_\beta}\,a\indices{^I_J_\alpha}\right)\\[\medskipamount]\fl
&+\onehalf\tilde{P}\indices{^J_K^\alpha^\beta}\left(\pfrac{A\indices{^K_J_\alpha}}{x^{\beta}}+\pfrac{A\indices{^K_J_\beta}}{x^{\alpha}}\right)-
\onehalf\tilde{p}\indices{^J_K^\alpha^\beta}\left(\pfrac{a\indices{^K_J_\alpha}}{x^{\beta}}+\pfrac{a\indices{^K_J_\beta}}{x^{\alpha}}\right).
\end{eqnarray*}
Defining a second amended Hamiltonian $\tilde{\HC}_{2}$ as
\begin{eqnarray}\fl
\tilde{\HC}_{2}=&\tilde{\HC}_0+
\rmi q\left(\tilde{\bar{\pi}}_{K}^{\beta}\phi^{J}-
\bar{\phi}_{K}\tilde{\pi}^{J\beta}\right)a\indices{^K_J_\beta}\nonumber\\\fl
&\mbox{}-\onehalf\rmi q\,\tilde{p}\indices{^J_K^\alpha^\beta}
\left(a\indices{^K_I_\alpha}\,a\indices{^I_J_\beta}-a\indices{^K_I_\beta}\,a\indices{^I_J_\alpha}\right)+
\onehalf\tilde{p}\indices{^J_K^\alpha^\beta}\left(\pfrac{a\indices{^K_J_\alpha}}{x^{\beta}}+\pfrac{a\indices{^K_J_\beta}}{x^{\alpha}}\right),\label{H-conv-gt}
\end{eqnarray}
this Hamiltonian yields the transformed second amended Hamiltonian $\tilde{\HC}_{2}^\prime$
\begin{eqnarray*}\fl
\tilde{\HC}_{2}^{\prime}&=\tilde{\HC}_0+
\rmi q\left(\tilde{\bar{\Pi}}_{K}^{\beta}\Phi^{J}-
\bar{\Phi}_{K}\tilde{\Pi}^{J\beta}\right)A\indices{^K_J_\beta}\\\fl
&\mbox{}-\onehalf\rmi q\tilde{P}\indices{^J_K^\alpha^\beta}
\left(A\indices{^K_I_\alpha}\,A\indices{^I_J_\beta}-A\indices{^K_I_\beta}\,A\indices{^I_J_\alpha}\right)+
\onehalf\tilde{P}\indices{^J_K^\alpha^\beta}\left(\pfrac{A\indices{^K_J_\alpha}}{x^{\beta}}+\pfrac{A\indices{^K_J_\beta}}{x^{\alpha}}\right).
\end{eqnarray*}
The expression~(\ref{H-conv-gt}) has the required functional form under the transformation generated by Eq.~(\ref{extgen2-general-gt}).
Its symmetric-derivative term is an intermediate compensation term in the action identity.
It is not an ordinary derivative-free Hamiltonian term to be varied independently of the canonical kinetic term.
Equivalently, the compensating contribution is
\begin{eqnarray}\fl
\tilde{\HC}_2-\tilde{\HC}_0&=\rmi q\left(\tilde{\bar{\pi}}_{K}^{\beta}\phi^{J}
-\bar{\phi}_{K}\tilde{\pi}^{J\beta}\right)a\indices{^K_J_\beta}\label{H-ext-gt}\\\fl
&\thickspace \thickspace \mbox{}-\onehalf\rmi q\,\tilde{p}\indices{^J_K^\alpha^\beta}\Big(a\indices{^K_I_\alpha}\,a\indices{^I_J_\beta}-a\indices{^K_I_\beta}\,a\indices{^I_J_\alpha}\Big)
+\onehalf\tilde{p}\indices{^J_K^\alpha^\beta}\left(\pfrac{a\indices{^K_J_\alpha}}{x^{\beta}}+\pfrac{a\indices{^K_J_\beta}}{x^{\alpha}}\right)\!.\nonumber
\end{eqnarray}
Combining this contribution with the canonical kinetic term in Eq.~(\ref{action-int5}) gives
\begin{eqnarray}\fl
S=\int_V&\rmd^{4}x\left[\left(\pfrac{\bar{\phi}_{I}}{x^{\beta}}+\rmi q\,\bar{\phi}_{K}\,a\indices{^K_I_\beta}\right)\tilde{\pi}^{I\beta}
+\tilde{\bar{\pi}}_{I}^{\beta}\left(\pfrac{\phi^{I}}{x^{\beta}}-\rmi q\,a\indices{^I_J_\beta}\,\phi^J\right)\right.\label{action-int6}\\\fl
&\left.\mbox{}+\onehalf\tilde{p}\indices{^J_K^\alpha^\beta}\left(\pfrac{a\indices{^K_J_\alpha}}{x^\beta}-\pfrac{a\indices{^K_J_\beta}}{x^\alpha}
+\rmi q\left(a\indices{^K_I_\alpha}\,a\indices{^I_J_\beta}-a\indices{^K_I_\beta}\,a\indices{^I_J_\alpha}\right)\right)
-\tilde{\HC}_0\right]\!.\nonumber
\end{eqnarray}
For reference, the resulting locally gauge-invariant action is
\begin{eqnarray}\fl
S=\int_{V}\rmd^{4}x&\left[\left(\pfrac{\bar{\phi}_{I}}{x^{\beta}}+\rmi q\,\bar{\phi}_{K}\,a\indices{^K_I_\beta}\right)\tilde{\pi}^{I\beta}
+\tilde{\bar{\pi}}_{I}^{\beta}\left(\pfrac{\phi^{I}}{x^{\beta}}-\rmi q\,a\indices{^I_J_\beta}\,\phi^J\right)\right.\label{action-int7}\\\fl
&\left.\mbox{}+\onehalf\tilde{p}\indices{^J_K^\alpha^\beta}\left(\pfrac{a\indices{^K_J_\alpha}}{x^\beta}-\pfrac{a\indices{^K_J_\beta}}{x^\alpha}
+\rmi q\left(a\indices{^K_I_\alpha}\,a\indices{^I_J_\beta}-a\indices{^K_I_\beta}\,a\indices{^I_J_\alpha}\right)\right)-\tilde{\HC}_0\right].\nonumber
\end{eqnarray}
This is the result of a Yang-Mills~\cite{YM54} --- also referred to as a non-Abelian --- gauge theory.
If the index ranges $I,J,K$ reduce to $I=J=K=1$, the self-coupling terms $a\indices{^K_I_\alpha}\,a\indices{^I_J_\beta}-a\indices{^K_I_\beta}\,a\indices{^I_J_\alpha}$ cancel.
The formalism then reduces to an Abelian gauge theory.
Each term of the final integrand is a scalar density when the seed Hamiltonian is completed with the appropriate metric contractions and density factors.
The internal field strength is
\begin{displaymath}
F\indices{^K_J_\mu_\nu}
=\pfrac{a\indices{^K_J_\nu}}{x^\mu}
-\pfrac{a\indices{^K_J_\mu}}{x^\nu}
-\rmi q\left(a\indices{^K_I_\mu}a\indices{^I_J_\nu}
-a\indices{^K_I_\nu}a\indices{^I_J_\mu}\right),
\end{displaymath}
and it transforms as $F_{\mu\nu}'=UF_{\mu\nu}U^\dagger$.
Thus the gauge kinetic term in Eq.~(\ref{action-int7}) is $-\onehalf\operatorname{tr}(\tilde p^{\mu\nu}F_{\mu\nu})$.
The minus sign follows from the convention that $\tilde p^{\mu\nu}$ is conjugate to $\partial_\nu a_\mu$.

For the Yang--Mills model considered here, we now choose the reduced momentum space
\begin{displaymath}
\tilde p\indices{^J_K^{(\mu\nu)}}=0.
\end{displaymath}
The homogeneous momentum transformation preserves this constraint.
The generating-function rules above were derived on the unrestricted component space and induce a transformation on this reduced space.
The final action is varied with antisymmetric $\delta\tilde p^{\mu\nu}$.
On that space,
\begin{displaymath}
\pfrac{\FC_2^\mu}{\tilde P\indices{^J_K^{\alpha\nu}}}
=\onehalf\left(\delta^\mu_\nu A\indices{^K_J_\alpha}
-\delta^\mu_\alpha A\indices{^K_J_\nu}\right).
\end{displaymath}
This projected relation replaces the unrestricted Kronecker-delta rule.
Cancellation of the symmetric derivative term alone would not imply the vanishing of the symmetric momentum for an unspecified Hamiltonian.
A model retaining independent symmetric components would require its own symmetric-sector algebraic equations.

The exterior derivative in $F_{\mu\nu}$ contains no affine connection.
Accordingly, this kinetic term has no explicit affine-connection source.
It does not follow that the gauge field decouples from gravity, since the metric-dependent terms in $\tilde{\HC}_0$ generally have a nonzero metric variation.
Additional invariant interactions involving torsion or curvature are not excluded.
An independent curvature sector of the form $-\onehalf\tilde q_\eta{}^{\alpha\beta\mu}R^\eta{}_{\alpha\beta\mu}$
can be added without changing the internal transformation law, with its dynamics included in $\tilde{\HC}_0$.
The distinction between metric coupling and direct affine-connection coupling is essential and is not a statement
of a gravitational decoupling in Kaluza--Klein theory~\cite{kaluza21}.
\subsection{Canonical field equations for $\phi^I$ and $\tilde{\bpi}_I$}
Variation of Eq.~(\ref{action-int7}) gives
\begin{eqnarray}
\pfrac{\phi^I}{x^\mu}
&=\pfrac{\tilde{\HC}_0}{\tilde{\bar\pi}_I^\mu}
+\rmi q\,a\indices{^I_J_\mu}\phi^J,\nonumber\\
\pfrac{\tilde\pi^{I\alpha}}{x^\alpha}
&=-\pfrac{\tilde{\HC}_0}{\bar\phi_I}
+\rmi q\,a\indices{^I_J_\alpha}\tilde\pi^{J\alpha}.
\label{covariant-deri}
\end{eqnarray}
The complex-conjugate equations follow in the same way.
Defining $D_\mu\phi^I=\partial_\mu\phi^I-\rmi q\,a\indices{^I_J_\mu}\phi^J$ gives the usual internal gauge-covariant form.
For the density divergence, $D_\mu\tilde\pi^{I\mu}$ denotes $\partial_\mu\tilde\pi^{I\mu}-\rmi q\,a\indices{^I_J_\mu}\tilde\pi^{J\mu}$.
This internal derivative is distinct from an affine covariant derivative.

We use the convention that the derivative index of $\gamma^\rho{}_{\sigma\mu}$ is the last index.
For a contravariant vector density of weight one,
\begin{displaymath}
\nabla_\mu\tilde V^\nu
=\partial_\mu\tilde V^\nu
+\gamma^\nu{}_{\rho\mu}\tilde V^\rho
-\gamma^\rho{}_{\rho\mu}\tilde V^\nu.
\end{displaymath}
Consequently, with $S^\rho{}_{\mu\nu}=\onehalf(\gamma^\rho{}_{\mu\nu}-\gamma^\rho{}_{\nu\mu})$,
\begin{displaymath}
\nabla_\alpha\tilde\pi^{I\alpha}
=\partial_\alpha\tilde\pi^{I\alpha}
+2S^\alpha{}_{\xi\alpha}\tilde\pi^{I\xi}.
\end{displaymath}
The ordinary divergence is already a scalar density under spacetime coordinate changes.
The torsion trace in this rewriting is not an additional gauge-field interaction in the action.
\subsection{Canonical field equations for $\ba\indices{^L_M}$ and $\tilde{\bp}\indices{^L_M}$}
On the reduced antisymmetric momentum space, variation of Eq.~(\ref{action-int7}) yields
\begin{eqnarray}
2\pfrac{\tilde{\HC}_0}{\tilde p\indices{^M_L^\nu^\mu}}
&=\pfrac{a\indices{^L_M_\nu}}{x^\mu}
-\pfrac{a\indices{^L_M_\mu}}{x^\nu}\nonumber\\
&\quad+\rmi q\left(a\indices{^L_I_\nu}a\indices{^I_M_\mu}
-a\indices{^L_I_\mu}a\indices{^I_M_\nu}\right)
=F\indices{^L_M_\mu_\nu}.
\label{can-momentum-gf-ext}
\end{eqnarray}
The Hamiltonian derivative is a projected derivative on the antisymmetric space.
Equation~(\ref{can-momentum-gf-ext}) is a tensor equation and does not itself prove antisymmetry of an unrestricted momentum.

Variation with respect to the gauge field gives
\begin{eqnarray}
\pfrac{\tilde p\indices{^M_N^\mu^\alpha}}{x^\alpha}
&=-\pfrac{\tilde{\HC}_0}{a\indices{^N_M_\mu}}
-\rmi q\left(\tilde{\bar\pi}_N^\mu\phi^M-\bar\phi_N\tilde\pi^{M\mu}\right)\nonumber\\
&\quad+\rmi q\left(\tilde p\indices{^M_J^\alpha^\mu}a\indices{^J_N_\alpha}
-a\indices{^M_J_\alpha}\tilde p\indices{^J_N^\alpha^\mu}\right).
\label{div-p-gf-ext}
\end{eqnarray}
The first term vanishes for a seed Hamiltonian independent of the gauge potential.
Using the affine derivative of a weight-one rank-two tensor density, the equivalent equation is
\begin{eqnarray}
\nabla_\alpha\tilde p\indices{^M_N^\mu^\alpha}
&=-\pfrac{\tilde{\HC}_0}{a\indices{^N_M_\mu}}
-\rmi q\left(\tilde{\bar\pi}_N^\mu\phi^M-\bar\phi_N\tilde\pi^{M\mu}\right)\nonumber\\
&\quad+\rmi q\left(\tilde p\indices{^M_J^\alpha^\mu}a\indices{^J_N_\alpha}
-a\indices{^M_J_\alpha}\tilde p\indices{^J_N^\alpha^\mu}\right)\nonumber\\
&\quad+\tilde p\indices{^M_N^\xi^\alpha}S^\mu{}_{\xi\alpha}
+2\tilde p\indices{^M_N^\mu^\xi}S^\alpha{}_{\xi\alpha}.
\label{div-p-gf-ext1}
\end{eqnarray}
The connection terms cancel when this equation is returned to the ordinary density-divergence form~(\ref{div-p-gf-ext}).
They do not represent the direct vector--torsion interaction considered in the next section.
\section{Diffeomorphism transformation in the extended canonical formalism\label{sec:diffeo-ext}}
We now consider ordinary vector fields without imposing the inhomogeneous internal U$(N)$ transformation law of \sref{sec:gen-gauge-ext}.
The labels $I,J,K$ are retained as internal component labels, but the vector field in this section is not the internal gauge connection of the preceding section.
This is a second application of the formalism, not an additional gauging of the same matter system.
We initially use unrestricted vector and connection momenta in the generating-function calculation.
The starting first-order action in the $x$ chart is
\begin{eqnarray}
S=\int_V\rmd^4x\Bigg[
\tilde{\bar\pi}_I^\beta\pfrac{\phi^I}{x^\beta}
+\pfrac{\bar\phi_I}{x^\beta}\tilde\pi^{I\beta}
+\tilde p\indices{^J_K^\alpha^\beta}\pfrac{a\indices{^K_J_\alpha}}{x^\beta}
+\tilde q\indices{_\eta^{\alpha\beta\mu}}\pfrac{\gamma^\eta{}_{\alpha\beta}}{x^\mu}
-\tilde{\HC}_0\Bigg].
\label{action-int8}
\end{eqnarray}
The symbol $\tilde{\HC}_0$ now includes the chosen invariant matter and geometric dynamics of this second model.
The same notation does not require it to be the Hamiltonian of \sref{sec:gen-gauge-ext}.
Any metric entering its contractions is transformed as a tensor.
It may be a prescribed background or an independent variational field, as specified by the chosen model.
No metric-compatibility or torsion-free condition is assumed.

The transformation rule under changes of the space-time location for the \emph{affine connection} $\gamma\indices{^{\xi}_{\alpha\beta}}$ is given by
\begin{equation}\label{Chris-trans}
\Gamma\indices{^{\eta}_{\mu\nu}}(\bX)=\gamma\indices{^{\xi}_{\kappa\lambda}}(\bx)\,\pfrac{X^{\eta}}{x^{\xi}}
\pfrac{x^{\kappa}}{X^{\mu}}\,\pfrac{x^{\lambda}}{X^{\nu}}+\pfrac{X^{\eta}}{x^{\xi}}\,\ppfrac{x^{\xi}}{X^{\mu}}{X^{\nu}}.
\end{equation}
The independent field $\gamma^\eta{}_{\alpha\beta}$ already has a conjugate momentum in Eq.~(\ref{action-int8}).
Its transformation law alone does not specify its dynamics.
The purpose of the following construction is to identify the compensating terms that turn the kinetic expression into a covariant action.
A gravitational dynamics Hamiltonian must still be supplied in $\tilde{\HC}_0$.

The map $x=x(X)$ is prescribed, as in Eq.~(\ref{genF2-ext}).
Its first, second, and third derivatives are explicit coordinate-dependent coefficients, not derivatives of a varied embedding field.
We continue to use the convention $\nabla_\beta a_\alpha=\partial_\beta a_\alpha-\gamma^\xi{}_{\alpha\beta}a_\xi$.
The curvature convention is
\begin{eqnarray*}
R^\eta{}_{\alpha\beta\mu}
&=\partial_\beta\gamma^\eta{}_{\alpha\mu}
-\partial_\mu\gamma^\eta{}_{\alpha\beta}
+\gamma^\eta{}_{\xi\beta}\gamma^\xi{}_{\alpha\mu}
-\gamma^\eta{}_{\xi\mu}\gamma^\xi{}_{\alpha\beta}.
\end{eqnarray*}
The extended generating function $\FC_{2}^{\mu}\big|_X$ that describes the transformation of the affine connection~(\ref{Chris-trans})
due to change of the reference frame, i.e.\ a chart transition is given by
\begin{eqnarray}
\FC_{2}^{\mu}\Big|_X&=\tilde{\bar{\Pi}}_{J}^{\mu}\,\phi^{J}+
\bar{\phi}_{J}\,\tilde{\Pi}^{J\mu}+\tilde{P}\indices{^J_K^\alpha^\mu}\,a\indices{^K_J_\beta}\pfrac{x^\beta}{X^\alpha}
\nonumber\\
&\quad+\tilde{Q}\indices{_{\eta}^{\alpha\beta\mu}}\left(\gamma\indices{^{\xi}_{\kappa\lambda}}\,
\pfrac{X^{\eta}}{x^{\xi}}\pfrac{x^{\kappa}}{X^{\alpha}}\pfrac{x^{\lambda}}{X^{\beta}}
+\pfrac{X^{\eta}}{x^{\xi}}\ppfrac{x^{\xi}}{X^{\alpha}}{X^{\beta}}\right).
\label{extgen3-general-gt}
\end{eqnarray}
This means that the affine connection $\gamma\indices{^{\eta}_{\alpha\beta}}(x)$ is now treated as a \emph{canonical variable} of the original system.
The new dynamical variable $\tilde{Q}\indices{_{\eta}^{\alpha\beta\mu}}(X)$ is formally introduced to represent its canonical conjugate at the transformed event $X$.
The transformation rules derived from the generating function (\ref{extgen3-general-gt}) are
\begin{eqnarray*}
\tilde{\bar{\pi}}_{J}^{\mu}(x)&=\pfrac{\FC_{2}^{\alpha}}{\phi^{J}}\pfrac{x^\mu}{X^\alpha}\detfrac{X}{x}=
\tilde{\bar{\Pi}}_{J}^{\alpha}(X)\,\pfrac{x^\mu}{X^\alpha}\detfrac{X}{x}\\
\bar{\Phi}_{J}\delta_{\nu}^{\mu}&=\pfrac{\FC_{2}^{\mu}}{\tilde{\Pi}^{J\nu}}=
\bar{\phi}_{J}\,\delta_{\nu}^{\mu}\\
\tilde{\pi}^{J\mu}(x)&=\pfrac{\FC_{2}^{\alpha}}{\bar{\phi}_{J}}\pfrac{x^\mu}{X^\alpha}\detfrac{X}{x}=
\tilde{\Pi}^{J\alpha}(X)\pfrac{x^\mu}{X^\alpha}\detfrac{X}{x}\\
\Phi^{J}\delta_{\nu}^{\mu}&=\pfrac{\FC_{2}^{\mu}}{\tilde{\bar{\Pi}}_{J}^{\nu}}=\phi^{J}\,\delta_{\nu}^{\mu}\\
\delta_{\nu}^{\mu}A\indices{^K_J_\alpha}(X)&=\pfrac{\FC_{2}^{\mu}}{\tilde{P}\indices{^J_K^\alpha^\nu}}=
\delta_{\nu}^{\mu}\,a\indices{^K_J_\beta}(x)\,\pfrac{x^\beta}{X^\alpha}\\
\tilde{p}\indices{^J_K^\beta^\mu}(x)&=\pfrac{\FC_{2}^{\xi}}{a\indices{^K_J_\beta}}\pfrac{x^\mu}{X^\xi}\detfrac{X}{x}=
\tilde{P}\indices{^J_K^\alpha^\xi}(X)\,\pfrac{x^\beta}{X^\alpha}\pfrac{x^\mu}{X^\xi}\detfrac{X}{x}\\
\delta_{\nu}^{\mu}\,\Gamma\indices{^{\eta}_{\alpha\beta}}(X)&=
\pfrac{\FC_{2}^{\mu}}{\tilde{Q}\indices{_{\eta}^{\alpha\beta\nu}}}=
\delta_{\nu}^{\mu}\left(\gamma\indices{^{\xi}_{\kappa\lambda}}(x)\,
\pfrac{X^{\eta}}{x^{\xi}}\pfrac{x^{\kappa}}{X^{\alpha}}\pfrac{x^{\lambda}}{X^{\beta}}+
\ppfrac{x^{\xi}}{X^{\alpha}}{X^{\beta}}\pfrac{X^{\eta}}{x^{\xi}}\right)\\
\tilde{q}\indices{_{\xi}^{\kappa\lambda\mu}}(x)&=
\pfrac{\FC_{2}^{\alpha}}{\gamma\indices{^{\xi}_{\kappa\lambda}}}\pfrac{x^\mu}{X^\alpha}\detfrac{X}{x}=
\tilde{Q}\indices{_{\eta}^{\alpha\beta\sigma}}(X)\,
\pfrac{X^{\eta}}{x^{\xi}}\pfrac{x^{\kappa}}{X^{\alpha}}
\pfrac{x^{\lambda}}{X^{\beta}}\pfrac{x^{\mu}}{X^{\sigma}}\detfrac{X}{x},
\end{eqnarray*}
Obviously, the desired rule for the connection coefficients from Eq.~(\ref{Chris-trans}) is reproduced.
The canonical transformation rule for the conjugate quantities $\tilde{q}\indices{_{\eta}^{\alpha\beta\mu}}$
represents the transformation rule for a mixed $(3,1)$-tensor density under an arbitrary change of the reference frame.
In contrast, the affine connection $\gamma\indices{^{\xi}_{\kappa\lambda}}$ does not transform as a tensor.

It remains to derive the transformation rule for the Hamiltonians.
According to Eq.~(\ref{genF2-ext}), it is obtained from the explicit $X$ divergence of the generator at fixed canonical fields.
All chart coefficients are differentiated.
\begin{eqnarray}\fl
\tilde{\HC}_3'&-\tilde{\HC}_3\detfrac{x}{X}
=\Hexpl{\FC_2^\mu}{X^\mu}\nonumber\\\fl
&=\tilde{P}\indices{^J_K^\alpha^\mu}\,a\indices{^K_J_\beta}\ppfrac{x^\beta}{X^\alpha}{X^\mu}\nonumber\\\fl
&\mbox{}+\tilde{Q}\indices{_{\eta}^{\alpha\beta\mu}}\!\!
\left[\gamma\indices{^{\xi}_{\kappa\lambda}}\!\left(\!
\ppfrac{X^{\eta}}{x^{\xi}}{x^{\nu}}\pfrac{x^{\nu}}{X^{\mu}}
\pfrac{x^{\kappa}}{X^{\alpha}}\pfrac{x^{\lambda}}{X^{\beta}}+
\ppfrac{x^{\kappa}}{X^{\alpha}}{X^{\mu}}
\pfrac{X^{\eta}}{x^{\xi}}\pfrac{x^{\lambda}}{X^{\beta}}+
\ppfrac{x^{\lambda}}{X^{\beta}}{X^{\mu}}
\pfrac{X^{\eta}}{x^{\xi}}\pfrac{x^{\kappa}}{X^{\alpha}}\!\right)\right.\nonumber\\\fl
&\qquad\qquad\quad\mbox{}+\left.
\pppfrac{x^{\xi}}{X^{\alpha}}{X^{\beta}}{X^{\mu}}
\pfrac{X^{\eta}}{x^{\xi}}+\ppfrac{x^{\xi}}{X^{\alpha}}{X^{\beta}}
\ppfrac{X^{\eta}}{x^{\xi}}{x^{\nu}}\pfrac{x^{\nu}}{X^{\mu}}\right]\!,
\label{pointtra-rules3-ext}
\end{eqnarray}
In order to replace the second derivatives of $x$, we make use of the transformation rule~(\ref{Chris-trans}) for the connection coefficients in the forms of
\begin{eqnarray*}
\ppfrac{x^{\kappa}}{X^{\alpha}}{X^{\beta}}&=\Gamma\indices{^{\tau}_{\alpha\beta}}\,
\pfrac{x^{\kappa}}{X^{\tau}}-\gamma\indices{^{\kappa}_{\tau\sigma}}\,
\pfrac{x^{\tau}}{X^{\alpha}}\pfrac{x^{\sigma}}{X^{\beta}}\\
\ppfrac{X^{\eta}}{x^{\xi}}{x^{\nu}}\pfrac{x^{\nu}}{X^{\mu}}&=\gamma\indices{^{\tau}_{\xi\sigma}}\,
\pfrac{X^{\eta}}{x^{\tau}}\pfrac{x^{\sigma}}{X^{\mu}}-
\Gamma\indices{^{\eta}_{\tau\mu}}\pfrac{X^{\tau}}{x^{\xi}}
\end{eqnarray*}
The term associated with the vector field then takes on the equivalent form
\begin{eqnarray*}
\tilde{P}\indices{^J_K^\alpha^\mu}\,a\indices{^K_J_\beta}\ppfrac{x^\beta}{X^\alpha}{X^\mu}&=
\tilde{P}\indices{^J_K^\alpha^\mu}\,a\indices{^K_J_\beta}\left(\Gamma\indices{^{\tau}_{\alpha\mu}}\,
\pfrac{x^{\beta}}{X^{\tau}}-\gamma\indices{^{\beta}_{\tau\sigma}}\,
\pfrac{x^{\tau}}{X^{\alpha}}\pfrac{x^{\sigma}}{X^{\mu}}\right)\\
&=\tilde{P}\indices{^J_K^\alpha^\mu}\,A\indices{^K_J_\tau}\,\Gamma\indices{^{\tau}_{\alpha\mu}}
-\tilde{p}\indices{^J_K^\alpha^\mu}\,a\indices{^K_J_\tau}\,\gamma\indices{^{\tau}_{\alpha\mu}}\detfrac{x}{X}.
\end{eqnarray*}
We observe that the vector field portion of transformation rule for the Hamiltonian no longer contains any
coefficient but could be expressed in a symmetric form in terms of the dynamic variables.

Proceeding to the terms proportional to $\tilde{Q}\indices{_{\eta}^{\alpha\beta\mu}}$, we first treat the
skew-symmetric portion of the tensor $\tilde{Q}\indices{_{\eta}^{\alpha\beta\mu}}$ in its indices $\beta$ and $\mu$.
All terms that are symmetric in $\beta,\mu$ then vanish, in particular the third order derivative term.
The entire expression can finally be expressed only in terms of the affine connection
\begin{eqnarray*}\fl
&\quad\tilde{Q}\indices{_{\eta}^{\alpha[\beta\mu]}}
\left[\gamma\indices{^{\xi}_{\kappa\lambda}}\!
\left(\ppfrac{X^{\eta}}{x^{\xi}}{x^{\nu}}
\pfrac{x^{\nu}}{X^{\mu}}\pfrac{x^{\kappa}}{X^{\alpha}}
\pfrac{x^{\lambda}}{X^{\beta}}+
\ppfrac{x^{\kappa}}{X^{\alpha}}{X^{\mu}}
\pfrac{X^{\eta}}{x^{\xi}}\pfrac{x^{\lambda}}{X^{\beta}}+
\cancel{\ppfrac{x^{\lambda}}{X^{\beta}}{X^{\mu}}
\pfrac{X^{\eta}}{x^{\xi}}\pfrac{x^{\kappa}}{X^{\alpha}}}\right)\right.\\\fl
&\qquad\qquad\mbox{}+\left.
\cancel{\pppfrac{x^{\xi}}{X^{\alpha}}{X^{\beta}}{X^{\mu}}
\pfrac{X^{\eta}}{x^{\xi}}}+\ppfrac{x^{\kappa}}{X^{\alpha}}{X^{\beta}}
\ppfrac{X^{\eta}}{x^{\kappa}}{x^{\nu}}\pfrac{x^{\nu}}{X^{\mu}}\right]\\\fl
&=\tilde{Q}\indices{_{\eta}^{\alpha[\beta\mu]}}
\left[\gamma\indices{^{\xi}_{\kappa\lambda}}
\left\{\thickspace \thickspace \,\left(\gamma\indices{^{\tau}_{\xi\sigma}}\,
\pfrac{X^{\eta}}{x^{\tau}}\pfrac{x^{\sigma}}{X^{\mu}}-
\Gamma\indices{^{\eta}_{\tau\mu}}\pfrac{X^{\tau}}{x^{\xi}}\right)
\pfrac{x^{\kappa}}{X^{\alpha}}\pfrac{x^{\lambda}}{X^{\beta}}\right.\right.\\\fl
&\qquad\qquad\mbox{}+\left.\left(\Gamma\indices{^{\tau}_{\alpha\mu}}\,
\pfrac{x^{\kappa}}{X^{\tau}}-\gamma\indices{^{\kappa}_{\tau\sigma}}\,
\pfrac{x^{\tau}}{X^{\alpha}}\pfrac{x^{\sigma}}{X^{\mu}}\right)
\pfrac{X^{\eta}}{x^{\xi}}\pfrac{x^{\lambda}}{X^{\beta}}\right\}\\\fl
&\qquad\qquad\mbox{}+\left.\left(\Gamma\indices{^{\tau}_{\alpha\beta}}\,
\pfrac{x^{\kappa}}{X^{\tau}}-\gamma\indices{^{\kappa}_{\tau\sigma}}\,
\pfrac{x^{\tau}}{X^{\alpha}}\pfrac{x^{\sigma}}{X^{\beta}}\right)
\left(\gamma\indices{^{\xi}_{\kappa\zeta}}\,
\pfrac{X^{\eta}}{x^{\xi}}\pfrac{x^{\zeta}}{X^{\mu}}-
\Gamma\indices{^{\eta}_{\xi\mu}}\pfrac{X^{\xi}}{x^{\kappa}}\right)\right]\\\fl
&=\tilde{q}\indices{_{\eta}^{\alpha[\beta\mu]}}
\gamma\indices{^{\xi}_{\alpha\beta}}\,\gamma\indices{^{\eta}_{\xi\mu}}\detfrac{x}{X}-
\tilde{Q}\indices{_{\eta}^{\alpha[\beta\mu]}}\,
\Gamma\indices{^{\xi}_{\alpha\beta}}\,\Gamma\indices{^{\eta}_{\xi\mu}}.
\end{eqnarray*}
We now treat the \emph{symmetric} portion of the tensor $\tilde{Q}\indices{_{\eta}^{\alpha\beta\mu}}$ in its indices $\beta$ and $\mu$
\begin{eqnarray*}\fl
&\quad\tilde{Q}\indices{_{\eta}^{\alpha(\beta\mu)}}
\left[\gamma\indices{^{\xi}_{\kappa\lambda}}\!
\left(\ppfrac{X^{\eta}}{x^{\xi}}{x^{\nu}}
\pfrac{x^{\nu}}{X^{\mu}}\pfrac{x^{\kappa}}{X^{\alpha}}
\pfrac{x^{\lambda}}{X^{\beta}}+
\ppfrac{x^{\kappa}}{X^{\alpha}}{X^{\mu}}
\pfrac{X^{\eta}}{x^{\xi}}\pfrac{x^{\lambda}}{X^{\beta}}+
\ppfrac{x^{\lambda}}{X^{\beta}}{X^{\mu}}
\pfrac{X^{\eta}}{x^{\xi}}\pfrac{x^{\kappa}}{X^{\alpha}}\right)\right.\\\fl
&\quad\mbox{}+\left.
\pppfrac{x^{\xi}}{X^{\alpha}}{X^{\beta}}{X^{\mu}}
\pfrac{X^{\eta}}{x^{\xi}}+\ppfrac{x^{\xi}}{X^{\alpha}}{X^{\beta}}
\ppfrac{X^{\eta}}{x^{\xi}}{x^{\nu}}\pfrac{x^{\nu}}{X^{\mu}}\right]\\\fl
&=\tilde{Q}\indices{_{\eta}^{\alpha(\beta\mu)}}
\left(\pfrac{\Gamma\indices{^{\eta}_{\alpha\beta}}}{X^\mu}-\pfrac{\gamma\indices{^{\xi}_{\kappa\lambda}}}{x^\tau}\,
\pfrac{x^\tau}{X^\mu}\,\pfrac{X^{\eta}}{x^{\xi}}\pfrac{x^{\kappa}}{X^{\alpha}}\pfrac{x^{\lambda}}{X^{\beta}}\right)\\\fl
&=\tilde{Q}\indices{_{\eta}^{\alpha(\beta\mu)}}\pfrac{\Gamma\indices{^{\eta}_{\alpha\beta}}}{X^\mu}-
\tilde{q}\indices{_{\eta}^{\alpha(\beta\mu)}}\pfrac{\gamma\indices{^{\eta}_{\alpha\beta}}}{x^\mu}\detfrac{x}{X}.
\end{eqnarray*}
The explicit chart coefficients have thus been eliminated in favor of field expressions and their derivatives.
The contributions have the same functional form and occur with opposite signs in the original and transformed variables.
The total transformation rule is
\begin{eqnarray*}
\tilde{\HC}_3^{\prime}-\tilde{\HC}_3\detfrac{x}{X}&=
\tilde{P}\indices{^J_K^\alpha^\mu}\,A\indices{^K_J_\tau}\,\Gamma\indices{^{\tau}_{\alpha\mu}}
-\tilde{p}\indices{^J_K^\alpha^\mu}\,a\indices{^K_J_\tau}\,\gamma\indices{^{\tau}_{\alpha\mu}}\detfrac{x}{X}\\
&+\onehalf\tilde{Q}\indices{_{\eta}^{\alpha\beta\mu}}\left(\pfrac{\Gamma\indices{^{\eta}_{\alpha\beta}}}{X^\mu}+\pfrac{\Gamma\indices{^{\eta}_{\alpha\mu}}}{X^\beta}
+\Gamma\indices{^{\xi}_{\alpha\mu}}\,\Gamma\indices{^{\eta}_{\xi\beta}}-\Gamma\indices{^{\xi}_{\alpha\beta}}\,\Gamma\indices{^{\eta}_{\xi\mu}}\right)\\
&-\onehalf\tilde{q}\indices{_{\eta}^{\alpha\beta\mu}}\!\left(\pfrac{\gamma\indices{^{\eta}_{\alpha\beta}}}{x^\mu}+\pfrac{\gamma\indices{^{\eta}_{\alpha\mu}}}{x^\beta}
+\gamma\indices{^{\xi}_{\alpha\mu}}\,\gamma\indices{^{\eta}_{\xi\beta}}-\gamma\indices{^{\xi}_{\alpha\beta}}\,\gamma\indices{^{\eta}_{\xi\mu}}\!\right)\!\detfrac{x}{X}\!.
\end{eqnarray*}
Thus a compensating expression with the required functional form is
\begin{equation}\label{H3-gauge-invariant}\fl
\tilde{\HC}_{3}=\tilde{\HC}_0+\tilde{p}\indices{^J_K^\alpha^\mu}\,a\indices{^K_J_\tau}\,\gamma\indices{^{\tau}_{\alpha\mu}}
+\onehalf\tilde{q}\indices{_{\eta}^{\alpha\beta\mu}}\left(\pfrac{\gamma\indices{^{\eta}_{\alpha\beta}}}{x^\mu}
+\pfrac{\gamma\indices{^{\eta}_{\alpha\mu}}}{x^\beta}+\gamma\indices{^{\xi}_{\alpha\mu}}\,\gamma\indices{^{\eta}_{\xi\beta}}
-\gamma\indices{^{\xi}_{\alpha\beta}}\,\gamma\indices{^{\eta}_{\xi\mu}}\right).
\end{equation}
Its difference from the invariant seed density is
\begin{equation}\label{H3-ext-gt-gen}\fl
\tilde{\HC}_{3}-\tilde{\HC}_0=\tilde{p}\indices{^J_K^\alpha^\mu}\,a\indices{^K_J_\tau}\,\gamma\indices{^{\tau}_{\alpha\mu}}
+\onehalf\tilde{q}\indices{_{\eta}^{\alpha\beta\mu}}\left(\pfrac{\gamma\indices{^{\eta}_{\alpha\beta}}}{x^\mu}+\pfrac{\gamma\indices{^{\eta}_{\alpha\mu}}}{x^\beta}
+\gamma\indices{^{\xi}_{\alpha\mu}}\,\gamma\indices{^{\eta}_{\xi\beta}}-\gamma\indices{^{\xi}_{\alpha\beta}}\,\gamma\indices{^{\eta}_{\xi\mu}}\right).
\end{equation}
The derivative-containing terms in Eqs.~(\ref{H3-gauge-invariant}) and~(\ref{H3-ext-gt-gen}) are intermediate compensation terms.
They must be combined with the canonical kinetic term before the equations of motion are varied.
Substitution into Eq.~(\ref{action-int8}) gives
\begin{eqnarray}\fl
S=\int_{V}\rmd^{4}x&\left[\pfrac{\bar{\phi}_{I}}{x^{\beta}}\tilde{\pi}^{I\beta}+\tilde{\bar{\pi}}_{I}^{\beta}\pfrac{\phi^{I}}{x^{\beta}}
+\tilde{p}\indices{^J_K^\alpha^\beta}\left(\pfrac{a\indices{^K_J_\alpha}}{x^\beta}
-a\indices{^K_J_\xi}\,\gamma\indices{^{\xi}_{\alpha\beta}}\right)\right.\nonumber\\\fl
&\left.\mbox{}+\onehalf\tilde{q}\indices{_{\eta}^{\alpha\beta\mu}}\left(\pfrac{\gamma\indices{^{\eta}_{\alpha\beta}}}{x^\mu}
-\pfrac{\gamma\indices{^{\eta}_{\alpha\mu}}}{x^\beta}
-\gamma\indices{^{\xi}_{\alpha\mu}}\,\gamma\indices{^{\eta}_{\xi\beta}}+\gamma\indices{^{\xi}_{\alpha\beta}}\,\gamma\indices{^{\eta}_{\xi\mu}}\right)
-\tilde{\HC}_0\right]\!.
\label{action-int9}
\end{eqnarray}
The chosen compensation produces a direct coupling of the ordinary vector field to the affine connection.
The vector derivative becomes $\nabla_\beta a\indices{^K_J_\alpha}$, while the connection terms combine to
$-\onehalf\tilde q_\eta{}^{\alpha\beta\mu}R^\eta{}_{\alpha\beta\mu}$ with the stated curvature convention.
The action~(\ref{action-int9}) thus writes equivalently in concise form
\begin{eqnarray}
S=\!\!\int_{V}\!\rmd^{4}x\!\left[\pfrac{\bar{\phi}_{I}}{x^{\beta}}\tilde{\pi}^{I\beta}+\tilde{\bar{\pi}}_{I}^{\beta}\pfrac{\phi^{I}}{x^{\beta}}
+\tilde{p}\indices{^J_K^\alpha^\beta}\,\nabla_\beta a\indices{^K_J_\alpha}
-\onehalf\tilde{q}\indices{_{\eta}^{\alpha\beta\mu}}\,R\indices{^\eta_{\alpha\beta\mu}}
-\tilde{\HC}_0\right]\!.\nonumber\\
\label{action-int10}
\end{eqnarray}
The integrand is a scalar density provided that $\tilde{\HC}_0$ is an invariant scalar density.
The metric and the connection need not be compatible, so $R^\eta{}_{\alpha\beta\mu}$ is the curvature of a general affine connection.
Only its antisymmetry in $\beta,\mu$ is used here.

For the curvature-sector model considered below, we now choose
\begin{displaymath}
\tilde q_\eta{}^{\alpha\beta\mu}
=\tilde q_\eta{}^{\alpha[\beta\mu]}.
\end{displaymath}
The homogeneous momentum transformation preserves this restriction.
The preceding generating-function calculation used unrestricted components, while the final action is varied with antisymmetric $\delta\tilde q_\eta{}^{\alpha\beta\mu}$.
The projected generator derivative is
\begin{displaymath}
\pfrac{\FC_2^\mu}{\tilde Q_\eta{}^{\alpha\beta\nu}}
=\onehalf\left(\delta^\mu_\nu\Gamma^\eta{}_{\alpha\beta}
-\delta^\mu_\beta\Gamma^\eta{}_{\alpha\nu}\right).
\end{displaymath}
The absence of the symmetric part from the curvature contraction does not prove that part to vanish for an arbitrary Hamiltonian.
Keeping an independent symmetric sector instead would require its own algebraic field equation.

Depending on the chosen ordinary-vector model, the momentum $\tilde{p}\indices{^J_K^\alpha^\beta}$ may be skew-symmetric in $\alpha$ and $\beta$.
The action~(\ref{action-int9}) then takes on the form
\begin{eqnarray}\fl
S=\!\int_{V}\rmd^{4}x\Bigg[\pfrac{\bar{\phi}_{I}}{x^{\beta}}\tilde{\pi}^{I\beta}+\tilde{\bar{\pi}}_{I}^{\beta}\pfrac{\phi^{I}}{x^{\beta}}
+\onehalf\tilde{p}\indices{^J_K^\alpha^\beta}\!\left(\nabla_\beta a\indices{^K_J_\alpha}-\nabla_\alpha a\indices{^K_J_\beta}\right)
-\onehalf\tilde{q}\indices{_{\eta}^{\alpha\beta\mu}}\,R\indices{^\eta_{\alpha\beta\mu}}
-\tilde{\HC}_0\Bigg]\!.\nonumber\\\label{action-int11}
\end{eqnarray}
Introducing the exterior-derivative tensor $f\indices{^K_J_\beta_\alpha}$, the action equivalently becomes
\begin{eqnarray*}
S=\int_{V}\rmd^{4}x&\Bigg[\pfrac{\bar{\phi}_{I}}{x^{\beta}}\tilde{\pi}^{I\beta}+\tilde{\bar{\pi}}_{I}^{\beta}\pfrac{\phi^{I}}{x^{\beta}}
+\onehalf\tilde{p}\indices{^J_K^\alpha^\beta}\,f\indices{^K_J_\beta_\alpha}\nonumber\\
&\thickspace \thickspace \mbox{}-\tilde{p}\indices{^J_K^\alpha^\beta}\,a\indices{^K_J_\xi}\,S\indices{^{\xi}_{\alpha\beta}}
-\onehalf\tilde{q}\indices{_{\eta}^{\alpha\beta\mu}}\,R\indices{^\eta_{\alpha\beta\mu}}
-\tilde{\HC}_0\Bigg],
\end{eqnarray*}
The torsion tensor and the exterior-derivative tensor are defined by
\begin{equation}\label{def-torsion_tensor}
S\indices{^{\xi}_{\alpha\beta}}=\onehalf\left(\gamma\indices{^{\xi}_{\alpha\beta}}-\gamma\indices{^{\xi}_{\beta\alpha}}\right),\qquad
f\indices{^K_J_\beta_\alpha}=\pfrac{a\indices{^K_J_\alpha}}{x^\beta}-\pfrac{a\indices{^K_J_\beta}}{x^\alpha}.
\end{equation}
For the covariant-derivative model adopted here, an antisymmetric vector momentum therefore gives a direct torsion interaction.
The choice is admissible for ordinary vector fields, including Proca-type models, but it is not forced by diffeomorphism invariance.
Indeed,
\begin{displaymath}
\tilde p^{\alpha\beta}a_\xi\gamma^\xi{}_{\alpha\beta}
=\tilde p^{(\alpha\beta)}a_\xi\gamma^\xi{}_{(\alpha\beta)}
+\tilde p^{[\alpha\beta]}a_\xi S^\xi{}_{\alpha\beta}.
\end{displaymath}
The inhomogeneous connection transformation is symmetric in the two lower indices.
Its compensation fixes the first contribution but not the second, which is already tensorial.
For an antisymmetric momentum, each member of the family
\begin{displaymath}
\tilde{\LC}_{\mathrm{kin}}^{(c)}
=\onehalf\tilde p^{\alpha\beta}
\left(\partial_\beta a_\alpha-\partial_\alpha a_\beta\right)
-c\,\tilde p^{\alpha\beta}a_\xi S^\xi{}_{\alpha\beta},
\qquad c\in\RB,
\end{displaymath}
is diffeomorphism covariant.
Internal component contractions are suppressed in this expression.
The choice $c=1$ gives Eq.~(\ref{action-int11}), while $c=0$ uses the exterior derivative without a direct torsion interaction.
The field equations below refer to the displayed $c=1$ model.

This interaction cannot simply be appended to the standard internal gauge system of \sref{sec:gen-gauge-ext}.
In the Abelian case, $a_\xi\mapsto a_\xi+\partial_\xi\chi$ changes it by
\begin{displaymath}
\delta_\chi\left(-\tilde p^{\alpha\beta}a_\xi S^\xi{}_{\alpha\beta}\right)
=-\tilde p^{\alpha\beta}S^\xi{}_{\alpha\beta}\partial_\xi\chi,
\end{displaymath}
which does not generally vanish or give a harmless boundary term.
The corresponding non-Abelian obstruction follows from the inhomogeneous part of Eq.~(\ref{gauge-tra1-ext}).
Other gauge-invariant torsion interactions are possible, but they require a separate invariant construction~\cite{delhom20}.
\subsection{Canonical field equations for $\ba\indices{^N_M}$ and $\tilde{\bp}\indices{^N_M}$}
For unrestricted vector momentum, variation of Eq.~(\ref{action-int10}) gives
\begin{equation}\label{can-momentum-gf-ext-3}
\pfrac{a\indices{^N_M_\nu}}{x^\mu}
=\pfrac{\tilde{\HC}_0}{\tilde p\indices{^M_N^\nu^\mu}}
+a\indices{^N_M_\tau}\gamma^\tau{}_{\nu\mu}.
\end{equation}
Equivalently,
\begin{equation}\label{can-momentum-gf-ext-3-cov}
\nabla_\mu a\indices{^N_M_\nu}
=\pfrac{\tilde{\HC}_0}{\tilde p\indices{^M_N^\nu^\mu}}.
\end{equation}
For a model defined on the antisymmetric momentum space, this equation is replaced by its antisymmetric projection rather than supplemented by an equation for the symmetric derivative.
Variation of Eq.~(\ref{action-int11}) yields
\begin{equation}\label{can-momentum-gf-ext-3-skew}
\onehalf f\indices{^N_M_{\mu\nu}}
-a\indices{^N_M_\tau}S^\tau{}_{\nu\mu}
=\pfrac{\tilde{\HC}_0}{\tilde p\indices{^M_N^\nu^\mu}}.
\end{equation}
The minus sign follows directly from $\nabla_\mu a_\nu=\partial_\mu a_\nu-\gamma^\tau{}_{\nu\mu}a_\tau$.

Variation with respect to the vector field gives
\begin{equation}\label{div-p-gf-ext-3}
\pfrac{\tilde p\indices{^M_N^\mu^\alpha}}{x^\alpha}
=-\pfrac{\tilde{\HC}_0}{a\indices{^N_M_\mu}}
-\tilde p\indices{^M_N^\tau^\alpha}\gamma^\mu{}_{\tau\alpha}.
\end{equation}
The equivalent density-covariant expression is
\begin{equation}\label{div-p-gf-ext-3-cov}
\nabla_\alpha\tilde p\indices{^M_N^\mu^\alpha}
-2\tilde p\indices{^M_N^\mu^\tau}S^\alpha{}_{\tau\alpha}
=-\pfrac{\tilde{\HC}_0}{a\indices{^N_M_\mu}}.
\end{equation}
This divergence equation holds both for unrestricted momentum and for the antisymmetric model with the coupling coefficient $c=1$.

\subsection{Canonical field equations for $\gamma^\mu{}_{\alpha\beta}$ and $\tilde q_\mu{}^{\alpha\beta\nu}$}
On the chosen antisymmetric connection-momentum space, variation of Eq.~(\ref{action-int10}) gives
\begin{eqnarray}
\pfrac{\tilde{\HC}_0}{\tilde q_\eta{}^{\alpha\beta\mu}}
&=\onehalf\left(
\partial_\mu\gamma^\eta{}_{\alpha\beta}
-\partial_\beta\gamma^\eta{}_{\alpha\mu}
+\gamma^\eta{}_{\xi\mu}\gamma^\xi{}_{\alpha\beta}
-\gamma^\eta{}_{\xi\beta}\gamma^\xi{}_{\alpha\mu}\right)\nonumber\\
&=-\onehalf R^\eta{}_{\alpha\beta\mu}.
\label{curvature-tensor}
\end{eqnarray}
The same curvature convention therefore appears in the action and in the momentum equation.
This relation fixes the momentum--curvature correlation only after the functional form of $\tilde{\HC}_0$ has been specified.

Variation with respect to the independent connection, with the boundary variation set to zero, gives
\begin{eqnarray*}
\partial_\alpha\tilde q_\kappa{}^{\tau\sigma\alpha}
&=\tilde q_\xi{}^{\tau\sigma\alpha}\gamma^\xi{}_{\kappa\alpha}
-\tilde q_\kappa{}^{\xi\sigma\alpha}\gamma^\tau{}_{\xi\alpha}
-\tilde p\indices{^J_K^\tau^\sigma}a\indices{^K_J_\kappa}
-\pfrac{\tilde{\HC}_0}{\gamma^\kappa{}_{\tau\sigma}}.
\end{eqnarray*}
The last Hamiltonian derivative must be retained when the invariant seed density has additional algebraic connection dependence, for example through torsion.
It vanishes for a seed density independent of the affine connection.
Using the induced covariant derivative of a tensor density of weight one gives
\begin{eqnarray}
\nabla_\alpha\tilde q_\kappa{}^{\tau\sigma\alpha}
-\tilde q_\kappa{}^{\tau\xi\alpha}S^\sigma{}_{\xi\alpha}
-2\tilde q_\kappa{}^{\tau\sigma\alpha}S^\xi{}_{\alpha\xi}
&=-\tilde p\indices{^J_K^\tau^\sigma}a\indices{^K_J_\kappa}
-\pfrac{\tilde{\HC}_0}{\gamma^\kappa{}_{\tau\sigma}}.
\label{curv-div-0}
\end{eqnarray}
Here $S^\sigma{}_{\xi\alpha}$ is the torsion tensor defined in Eq.~(\ref{def-torsion_tensor}).
The explicit vector source is a consequence of the chosen covariant-derivative interaction.
For the Yang--Mills action of \sref{sec:gen-gauge-ext}, that source is absent because its kinetic term contains no affine connection.
For the antisymmetric ordinary-vector family with coefficient $c$, the displayed vector source is multiplied by $c$.

If the metric is also varied independently and the displayed action has no metric-derivative term, its equation follows from the metric variation of $\tilde{\HC}_0$.
Additional metric-momentum or other geometric sectors would require their own canonical terms and would modify the corresponding source equations.
The affine transformation law does not impose $\nabla_\lambda g_{\mu\nu}=0$.
Metric compatibility, a torsion-free restriction, and a particular gravitational dynamics must be specified or derived from the chosen model rather than inferred from covariance alone.
\section{Conclusions}
The presented extended canonical formalism provides a parametrized description of covariant Hamiltonian field theory and a systematic framework for constructing compensating interactions.
Coordinate reparametrization is distinct from independent gravitational dynamics.
Canonical transformations preserve the action principle, while a symmetry additionally requires the transformed Hamiltonian to retain its functional form.

For the local U$(N)$ transformation of complex scalar fields, the construction yields the usual matter--gauge-field coupling and the Yang--Mills field strength.
The antisymmetric momentum space is specified explicitly and the final action is varied within that space.
With the standard exterior-derivative field strength, neither the scalar kinetic term nor the Yang--Mills kinetic term has an explicit affine-connection source.
Metric-dependent terms nevertheless couple these fields to gravity.

For ordinary vector fields without the inhomogeneous internal gauge transformation, the affine covariant-derivative prescription gives a direct connection coupling.
Its antisymmetric momentum sector contains a vector--torsion interaction.
This is a consistent coupling choice, not a uniqueness consequence of diffeomorphism covariance.
The coefficient of an interaction that is already a scalar density remains unconstrained by the compensation argument.
The same term is not invariant under the standard internal gauge transformation.

The curvature sector is consistently described by $-\onehalf\tilde q_\eta{}^{\alpha\beta\mu}R^\eta{}_{\alpha\beta\mu}$ and its associated projected momentum equation.
The connection source equation includes any additional connection dependence of the seed Hamiltonian.
No metric-compatibility condition, unique gravitational dynamics, or curvature-induced vector mass follows from the canonical transformation law alone.
Once the invariant dynamics Hamiltonian and any further constraints are specified, the resulting action supplies the appropriate field equations for that model.
Questions of propagation, stability, and physical predictions then depend on that additional choice.

\end{document}